\documentclass[11pt]{article}

\usepackage{graphicx}        
\usepackage{amsmath,amsfonts,amssymb}
\usepackage{lineno,hyperref}
\usepackage{cite}
\usepackage{color}

\usepackage[margin=1in]{geometry}

\newcommand{\bgqar}{\begin{eqnarray}}
\newcommand{\enqar}{\end{eqnarray}}
\newcommand{\bgqarn}{\begin{eqnarray*}}
\newcommand{\enqarn}{\end{eqnarray*}}
\newcommand{\bgary}{\begin{array}}
\newcommand{\enary}{\end{array}}
\newcommand{\bgitem}{\begin{itemize}}
\newcommand{\enitem}{\end{itemize}}
\newcommand{\bc}{\begin{center}}
\newcommand{\ec}{\end{center}}
\newcommand{\beq}{\begin{equation}}
\newcommand{\eeq}{\end{equation}}
\newcommand{\bld}[1]{\mbox{\boldmath $#1$}}
\newcommand{\be}{\begin{enumerate}}
\newcommand{\ee}{\end{enumerate}}
\newcommand{\bL}{\begin{Large}}
\newcommand{\eL}{\end{Large}}
\newcommand{\bl}{\begin{large}}
\newcommand{\el}{\end{large}}
\newcommand{\noin}{\noindent}

\newcommand{\hc}[1]{\hspace*{#1cm}}

\newcommand{\Blue}{\color{blue}}
\newcommand{\Red}{\color{red}}

\newcommand{\df}{:=}

\newcommand{\OLS}{\mbox{\scriptsize OLS}}
\newcommand{\WLS}{\mbox{\scriptsize WLS}}

\newcommand{\vk}{\bld{k}}

\title{A Stochastic Global Identification Framework for Aerospace Vehicles Operating Under Varying Flight States}

\author{Fotis Kopsaftopoulos\thanks{Corresponding author. Email: fkopsaf@stanford.edu.}, Raphael Nardari, Yu-Hung Li and Fu-Kuo Chang \\ {\small Department of Aeronautics and Astronautics, Stanford University, CA, USA }}

\begin{document}

\maketitle \thispagestyle{empty} 

\begin{abstract}
In this work, a novel data-based stochastic ``global'' identification framework is introduced for air vehicles operating under varying flight states and uncertainty. In this context, the term ``global'' refers to the identification of a model that is capable of representing the system dynamics under any admissible flight state based on data recorded from a sample of these states. The proposed framework is based on stochastic time-series models for representing the system dynamics and aeroelastic response under multiple flight states, with each state characterized by several variables, such as the airspeed, angle of attack, altitude, temperature, etc., forming a flight state vector. The method's cornerstone lies in the new class of Vector-dependent Functionally Pooled (VFP) models which allow the explicit analytical inclusion of the flight state vector into the model parameters and, hence, system dynamics. This is achieved via the use of functional data pooling techniques for optimally treating  --as a single entity-- the data records corresponding to the various flight states. In this proof-of-concept study the flight state vector is defined by two variables, namely the airspeed and angle of attack of the vehicle. The experimental evaluation and assessment is based on a prototype bio-inspired self-sensing composite wing that is subjected to a series of wind tunnel experiments under multiple flight states. Distributed micro-sensors in the form of stretchable sensor networks are embedded in the composite layup of the wing in order to provide the sensing capabilities. Experimental data collected from piezoelectric sensors are employed for the identification of a stochastic ``global'' VFP model via appropriate parameter estimation and model structure selection methods. The estimated VFP model parameters constitute two-dimensional functions of the flight state vector defined by the airspeed and angle of attack. The identified model is able to successfully represent the wing's aeroelastic response under the admissible flight states via a minimum number of estimated parameters compared to standard identification approaches. The obtained results demonstrate the high accuracy and effectiveness of the proposed global identification  framework, thus constituting a first step towards the next generation of ``fly-by-feel'' aerospace vehicles with state awareness capabilities.
\end{abstract}



\clearpage

\section*{Important conventions and symbols}

\noin Definition is indicated by $\df$. Matrix transposition is indicated by the superscript $T$.

\noin Bold-face upper/lower case symbols designate matrix/column-vector quantities, respectively.
 
\noin A functional argument in parentheses designates function of a real variable; for instance $P(x)$ is a function of the real variable $x$.

\noin A functional argument in brackets designates function of an integer variable; for instance $x[t]$ is a function of normalized discrete time $(t=1,2,\ldots)$. The conversion from discrete normalized time to analog time is based on $(t-1)T_s$, with $T_s$ designating the sampling period.

\noin A hat designates estimator/estimate; for instance $\widehat{\bld{\theta}}$ is an estimator/estimate of $\bld{\theta}$.


\section*{Acronyms}

\noin\begin{tabular}{lcl} 
AoA  & : & Angle of attack \\
AR   & : & Autoregressive  \\
ARMA & : & Autoregressive moving average \\
ARMAX  & : & Autoregressive moving average with exogenous excitation  \\
ARX  & : & Autoregressive with exogenous excitation  \\
BIC  & : & Bayesian information criterion  \\
CFD  & : & Computational fluid dynamics \\
CMOS & : & Complementary metal-oxide-semiconductor \\ 
FEM  & : & Finite element model \\
FP   & : & Functionally pooled  \\
FRF  & : & Frequency response function  \\
GA   & : & Genetic algorithm  \\
HALE & : & High altitude long endurance \\
iid  & : & identically independently distributed  \\
LCO  & : & Limit-cycle oscillation \\
LPV  & : & Linear parameter varying \\
MA   & : & Moving average \\
MEMS & : & Micro-electro-mechanical systems \\
NLS  & : & Nonlinear least squares  \\
OLS  & : & Ordinary least squares  \\
PCB  & : & Printed circuit board \\
PE   & : & Prediction error  \\
PZT  & : & Lead zirconate titanate \\
RSS  & : & Residual sum of squares  \\
RTD  & : & Resistive temperature detector \\
SACL & : & Structures and composites laboratory \\
SHM  & : & Structural health monitoring  \\
SQP  & : & Sequential quadratic programming  \\
SSS  & : & Signal sum of squares  \\
UAV  & : & Unmanned aerial vehicle \\
VFP  & : & Vector-dependent functionally pooled  \\
WLS  & : & Weighted least squares  \\
X    & : & Exogenous
\end{tabular} 

\newpage\pagebreak 

\tableofcontents 


\section{Introduction} \label{sec:intro}

The next generation of intelligent aerospace structures and air vehicles will be able to ``feel'', ``think'', and ``react'' in real time based on high-resolution state-sensing, awareness, and self-diagnostic capabilities. They will be able to sense and observe phenomena at unprecedented length and time scales allowing for real-time optimal control and decision making, significantly improved performance, adaptability, autonomous operation, increased safety, reduced mission and maintenance costs, and complete life-cycle monitoring and management. One of the main challenges of the current state-of-the-art research is the development of technologies that will lead to autonomous ``fly-by-feel'' air vehicles inspired by the unprecedented sensing and actuation capabilities of biological systems. Such intelligent air vehicles will be able to (i) sense the external environment (temperature, air pressure, humidity, etc.) \cite{Lanzara-etal10, Salowitz-etal14}, (ii) sense their flight and aeroelastic state (airspeed, angle of attack, flutter, stall, aerodynamic loads, etc.) and internal structural condition (stresses, strains, damage) \cite{Kopsaftopoulos-etal-IWSHM15, Kopsaftopoulos-etal-SPIE16}, and (iii) effectively interpret the sensing data to achieve real-time state awareness and health monitoring \cite{Ihn-Chang04a, Ihn-Chang04b, Ihn-Chang08, Janapati-etal16, Larrosa-etal14, Zhuang-etal-IWSHM15, Kopsaftopoulos-Fassois10, Kopsaftopoulos-Fassois13}. Towards this end, novel data-driven approaches are needed for the accurate interpretation of sensory data collected under varying flight states, structural conditions, and uncertainty in complex dynamic environments.

The most critical challenge for the postulation of a complete and applicable data-driven state-awareness framework for aerospace vehicles is the effective modeling and interpretation of sensory data obtained under constantly changing dynamic environments, multiple flight states and varying structural health conditions. Evidently, all these different operating conditions have a significant impact on the vehicle dynamics and aeroelastic response. When it comes to the aeroelastic behavior, dynamic aeroelastic effects resulting from the interaction of the aerodynamic, elastic, and inertial forces require careful consideration throughout the design phase of the air vehicle and pose a major safety-critical factor in the qualification of aircraft into service \cite{Livne03, Henshaw-etal07, Georgiou-etal12, Mardanpour-Hodges15, Jones-Cesnik16, Kitson-Cesnik2016}. Accurate modeling and prediction of the aeroelastic response is a complex and challenging task with the undesirable effects including significant vibrations (for passengers and/or crew), airframe fatigue, loss of control, degraded performance, or even complete destruction of the vehicle itself. It is therefore evident that the flight states and operating conditions --characterized by one or more measurable variables, such as the airspeed, angle of attack (AoA), altitude, temperature, and so on-- may vary over time, and consequently affect the system dynamics and aeroelastic response. In such cases, the problem of identifying a single ``global'' model of the system that is capable of representing the dynamics under any admissible operating condition and multiple equilibria points based on available response and/or excitation signals poses a major challenge that needs to be properly addressed.

In the context of aerospace structures, aeroelastic analysis and prediction, this challenge is typically tackled via the identification of a number of distinct models, via the use of acceleration or dynamic strain data, with each model corresponding to a \emph{single} flight state; one model is identified for each constant airspeed resulting to an array of models covering the required airspeed range. Usually, the models employed are time-series autoregressive moving average (ARMA) or state-space representations in the time or frequency domains \cite{Henshaw-etal07, Torii-Matsuzaki01, Mcnamara-Friedmann07,  Basseville-etal07, Zouari-etal12, Dowell-Hall01, DeCallafon-etal09, Hallissy-etal11, Razak-etal11, Huang-etal16, Pang-Cesnik16, Sodja-etal16, Suh-etal13, Suh-etal14, Zeng-etal12, Zeng-Kukreja13, Neu-etal16}, frequency-domain time-varying models with additional exogenous excitation within the bandwidth of interest \cite{Ertveldt-etal14, Heeg-Morelli11}, or Linear Parameter Varying (LPV) models \cite{Hjartarson-etal13, Lind-etal05, Baldelli-etal07, Ryan-etal14, Zeng-etal10, Matsuzaki11, Toth10}. The latter are dynamical models with parameters expressed as functions of the variable(s) --referred to as scheduling variable(s)-- that designate the operating condition. The LPV model identification is based on the so-called local approach \cite{Toth10} that splits the problem into two distinct subproblems: (i) first, a number of local (or else frozen) models --each corresponding to a single operating condition for which response signals are available-- are estimated using conventional identification techniques \cite{Ljung99, Soderstrom-Stoica89}, and (ii) second, the parameters of the identified models are interpolated in order to provide a single global model \cite{Toth10}. For example, a flutter suppression control system for the X-56\footnote{X-56 is modular experimental research aircraft designed by Lockheed Martin under contract for the Air Force Research Laboratory (AFRL).} aircraft was developed in \cite{Ryan-etal14} based on an LVP-based method utilizing data from seven flight conditions and corresponding state-space models with the aircraft velocity being the scheduling variable. In addition, an LPV-based method for data-based flutter modeling and prediction was developed in \cite{Baldelli-etal07} where the model parameters from 21 distinct flight conditions were modeled as a polynomial fit function of the dynamic pressure.

The LPV-based approach is a straightforward extension of classical identification. Yet, when viewed within a stochastic framework in which the response signals are random in nature (stochastic), it may lead to suboptimal accuracy in terms of parameter estimation and representation of the system dynamics. The intuitive explanation is based on the fact that the signals are not treated as a single data entity generated from the same dynamical system, but rather in complete isolation from each other within a seemingly unrelated context in the process of obtaining each local model. This not only does neglect potential cross-correlations among the signal pairs, thus resulting into loss of information, but additionally leads to an unnecessarily high number of estimated parameters, thus violating the principle of statistical parsimony \cite[p. 492]{Ljung99}. In addition, this may further lead to increased estimation variance and thus reduced accuracy (lack of efficiency in statistical terminology) \cite[pp. 560--562]{Ljung99}. Finally, additional loss of accuracy and potentially increased error is involved in the subsequent interpolation of the obtained local models when constructing the LPV (global) model. This identification process leads to a global, but suboptimal, system representation.

In an effort to effectively tackle the aforementioned challenges and provide an accurate, efficient, compact, and parsimonious global model, the  \underline{\textbf{aim}} of the present study is the introduction and experimental evaluation of a novel data-based stochastic ``global'' identification framework for aerospace vehicles operating in dynamic environments under varying flight states (multiple equilibria) and uncertainty. The proposed approach allows the accurate modeling of the vehicle dynamics and aeroelastic response via a single and compact time series model. It is based on the notion of functional data pooling and incorporates statistical parameter estimation and model structure selection techniques for representing the system dynamics under any admissible flight state based on data recorded from sample states. More specifically, in this study the novel class of Vector-dependent Functionally Pooled (VFP) models \cite{Kopsaftopoulos12, Kopsaftopoulos-Fassois12} is introduced for the first time within the context of flight state awareness and aeroelastic response\footnote{An application of the VFP model representation within the context of SHM and precise damage localization and quantification in continuous structural topologies may be found in \cite{Kopsaftopoulos-Fassois13, Kopsaftopoulos12, Sakaris-etal16}.}. 

The \emph{unique} characteristic of the VFP model structure is that the model parameters and residual covariance series (noise sequences) are explicit functions of the flight state that is defined by several variables; in this present study the flight state is defined by a vector consisting of the airspeed and AoA. This functional dependency, which is critical for the accurate modeling and constitutes the cornerstone of the proposed framework, is achieved via the projection of the VFP model parameters onto appropriately selected functional subspaces spanned by mutually independent basis functions of the flight state vector. The class of VFP models resembles that of LPV, with some critical differences: (i) the signals are treated as a single entity and potential cross-correlations are accounted for to increase the modeling accuracy and estimation efficiency, (ii) the number of estimated parameters is minimal compared to multi-model and LPV approaches (parsimonious representation), and (iii) the estimation is accomplished in a \emph{single} step (instead of two subsequent steps) for achieving optimal accuracy. The current limitation of the VFP representation (as well as of the LPV approach) is that the data records used in the identification process need to be recorded from flight states that are kept constant for the duration of the data collection (equilibrium state), that is a common practice for the aeroelastic modeling and analysis \cite{Hjartarson-etal13, Lind-etal05, Baldelli-etal07, Ryan-etal14, Zeng-etal10, Matsuzaki11}. Thus, in its current form the framework may be applied for the identification of aerospace systems with slow-varying dynamics (varying airspeed, AoA, altitude, temperature, etc.), such as highly flexible high-altitude long-endurance (HALE) air vehicles \cite{Jones-Cesnik16, Kitson-Cesnik2016, Hallissy-etal11, Babbar-etal13} and unmanned aerial vehicles (UAV) for various applications (aerial photography, reconnaissance operations, inspection of infrastructure, environment/forest/land/wildlife monitoring, humanitarian aid, etc). Air vehicles undergoing aggressive maneuvering and rapid changes in their attitude cannot be accurately modeled with the proposed approach. Appropriate extensions to fast-evolving systems with time-varying dynamics are the subject of ongoing work and will be presented in a future study.

The main novel aspects of this study include:
\begin{itemize}
\item[(a)] Introduction of a novel data-based stochastic ``global'' identification framework for modeling the aeroelastic response of air vehicles under varying flight states characterized by several variables. In this work the flight states are characterized by varying airspeed and AoA.
\item[(b)] Proof-of-concept experimental assessment based on an intelligent composite wing with embedded bio-inspired distributed sensor networks subjected to a series of 266 wind tunnel experiments under varying flight states.
\item[(c)] Experimental stochastic identification and accurate ``global'' modeling of the wing dynamics and aeroelastic response for all admissible flight states.
\item[(d)] Accurate identification and monitoring of the dynamic flutter and stall phenomena under varying flight states.
\item[(e)] Experimental statistical analysis of the piezoelectric signal energy under varying flight states and correlation with the aerodynamic stall for monitoring and early detection.
\end{itemize}

The rest of the paper is organized as follows: The problem statement is presented in Section \ref{sec:problem}. The stochastic global identification framework is presented in Section \ref{sec:VFP-models}. The bio-inspired stretchable sensor networks and the composite wing integration are briefly outlined in Section \ref{sec:wing-sensors}. The wind tunnel experiments are described in Section \ref{sec:experiments}, while the wind-tunnel experimental results and the discussion are presented in Section \ref{sec:results}. Finally, the conclusions and future work are summarized in Section \ref{sec:conclusions}.

\section{Problem Statement} \label{sec:problem}

\begin{figure}[t!]
    \centering
    \includegraphics[scale=0.55]{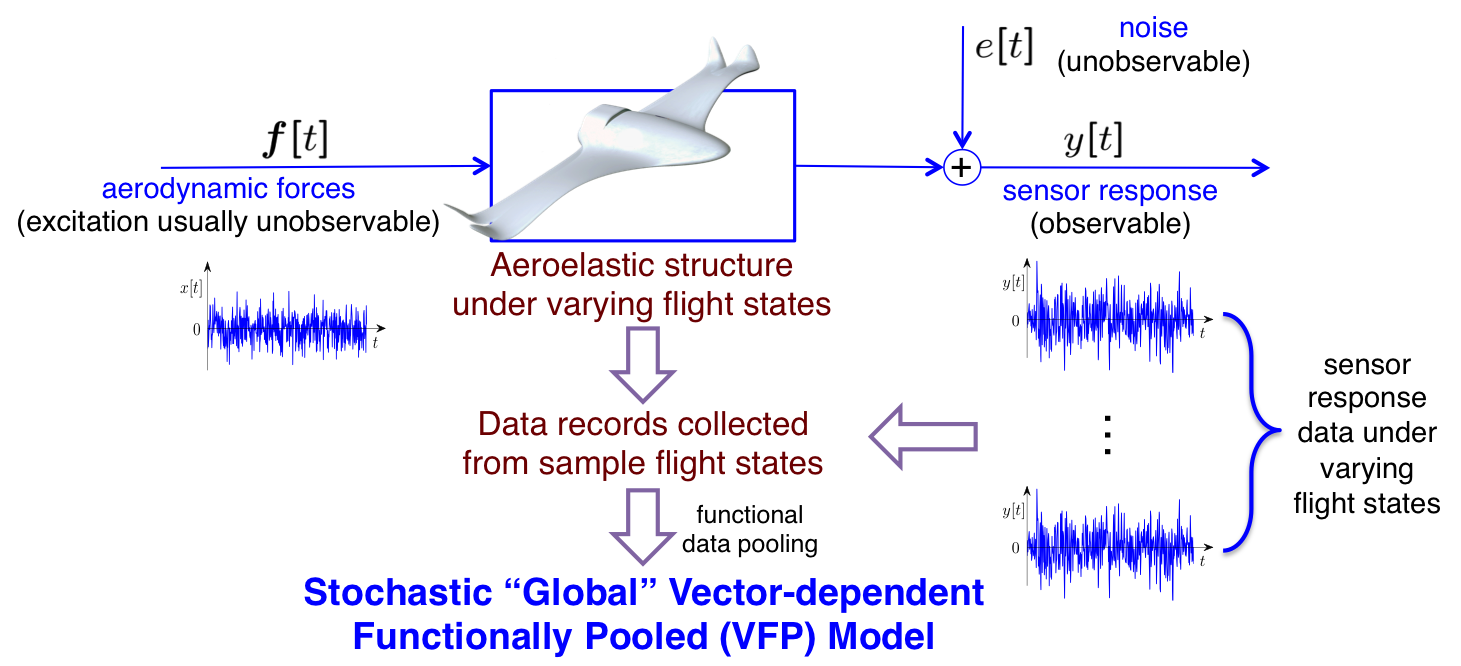}
    \caption{\label{fig:problem} \textbf{Stochastic ``global'' identification framework.} The stochastic global identification framework utilizing noise-corrupted data records obtained under a sample of all the admissible flight states for the estimation of a VFP model capable of representing the dynamics and aeroelastic response of the structure..}
\end{figure}

The problem statement of this work is as follows: Given dynamic noise-corrupted data records (response and/or excitation-response signal pairs) collected from a sample of the admissible flight states (multiple equilibria points), with each state characterized by several measurable variables (airspeed, AoA, altitude, temperature, etc.) and kept constant for the duration of the data collection, the goal is to identify a stochastic ``global'' model of the aerospace vehicle that is capable of accurately representing the aeroelastic response under uncertainty for all the admissible flight states. The stochastic model should be able to simultaneously treat the available data records as a single entity and account for data cross-correlation in order to make optimal use of the available information and achieve accurate parameter estimation. Special emphasis is placed on the successful modeling, monitoring, and representation of the critical aerodynamic phenomena of stall and flutter. Figure \ref{fig:problem} depicts a schematic representation of the proposed global identification framework.

In order to achieve the experimental evaluation and assessment of the proposed framework, a prototype proof-of-concept self-sensing composite UAV wing was designed and fabricated \cite{Kopsaftopoulos-etal-IWSHM15,Kopsaftopoulos-etal-SPIE16}. The wing is outfitted with bio-inspired stretchable sensor networks \cite{Lanzara-etal10, Salowitz-etal14, Salowitz-etal13, Salowitz-etal-IWSHM13, Guo-etal11, Guo14} consisting of distributed micro-sensors that enable its self-sensing capabilities. The sensor networks are embedded inside the composite layup of the wing, comprising carbon fiber and fiberglass plies, leaving a minimal parasitic footprint on the mechanical properties. In this work, piezoelectric sensors are used to sense the aeroelastic response (vibration) of the wing and allow the stochastic global identification of the dynamics and aeroelastic behavior under varying flight states, as well as the early detection of incipient dynamic flutter and stall. A series of 266 wind tunnel experiments, with each corresponding to a distinct AoA and airspeed (also referred to as freestream velocity) pair, are conducted for collecting data under a broad range of flight states. The experimental evaluation of the developed identification approach results in a single ``global'' VFP time-series model capable of accurately representing the wing dynamics across all the admissible flight states; that is all airspeeds and AoA considered in the wind tunnel experiments that may form the flight envelope of the vehicle.

\section{Stochastic Global Identification under Multiple Flight States} \label{sec:VFP-models}

In this section the ``global'' identification of the vehicle dynamics and aeroelastic response is addressed via the use of stochastic Functionally Pooled (FP) models, and specifically via the Vector-dependent Functionally Pooled AutoRegressive (VFP-AR) model structure \cite{Kopsaftopoulos12,Kopsaftopoulos-Fassois12}. These models are capable of representing the system dynamics for the complete range of flight states (airspeeds, AoA, altitudes, etc.) based on data records obtained under a sample of these states. The problem is important in a number of practical applications and is tackled within the recently introduced FP framework \cite{Kopsaftopoulos12,Kopsaftopoulos-Fassois12, Sakellariou-Fassois16}. This proof-of-concept study focuses on the case of flight states characterized by two variables, namely the airspeed and AoA of the wing. 

\subsection{Baseline modeling under a single flight state}

The baseline modeling under a single flight state is an initial step performed in order to facilitate --in the sense of providing approximate model orders-- the subsequent step of the global modeling under all admissible flight states.

A single data set under a specific flight state is obtained --either via actual flight testing, appropriately designed wind tunnel experiments, or calibrated high-fidelity computational aeroelastic models-- based on which an interval estimate of a discrete-time model (or a vector model or an array of models in the case of several response measurement locations) representing the system dynamics is obtained via standard identification procedures \cite{Ljung99,Soderstrom-Stoica89}. In this study, response-only AutoRegressive (AR) models are employed, as the wind tunnel airflow excitation signal is not measurable. However, in the case where the excitation signal can also be recorded via the use of appropriate sensors, the excitation-response AutoRegressive with eXogenous excitation (ARX) model structure may be a more appropriate representation that could potentially offer increased modeling accuracy \cite{Kopsaftopoulos-Fassois13}. Alternatively, depending on the properties and nature of the system dynamics, response signals and corrupting noise, more elaborate representations, such as the generic AutoRegressive Moving Average with eXogenous excitation (ARMAX) may be used \cite{Ljung99,Soderstrom-Stoica89,Box-etal94}. 

An AR$(n)$ model is of the following form \cite{Ljung99}:

\beq y[t] + \sum_{i=1}^{n} a_i \cdot y[t-i] =  e[t] \qquad e[t] \sim \, \mbox{iid} \, \mathcal{N} \bigl( 0,\sigma^2_e \bigr) \label{eq:ar-model} \eeq
with $t$ designating the normalized discrete time ($t=1,2,3,\ldots$ with absolute time being $(t-1) T_s$, where $T_s$ stands for the sampling period), $y[t]$ the measured vibration response signals as generated by the piezoelectric sensors of the wing, $n$ the AR order, and $e[t]$ the stochastic model residual (one-step-ahead prediction error) sequence, that is a white (serially uncorrelated), Gaussian, zero mean with variance $\sigma^2_e$ sequence. The symbol $\mathcal{N}(\cdot,\cdot)$ designates Gaussian distribution with the indicated mean and variance, and iid stands for identically independently distributed.

The model is parameterized in terms of the parameter vector $\bld{\bar{\theta}} = [a_1 \; \ldots \; a_n \; \vdots \; \sigma^2_e]^T$ to be estimated from the measured response signals \cite{Ljung99}. Model estimation may be achieved based on minimization of the Ordinary Least Squares (OLS) or Weighted Least Squares (WLS) criteria \cite{Ljung99}. The modeling procedure involves the successive fitting of AR$(n)$ models for increasing order $n$ until an adequate model is selected \cite{Fassois01}. Model order selection is based on the Bayesian Information Criterion (BIC) and the residual sum of squares normalized by the signal sum of squares (RSS/SSS). Final model validation is based on formal verification of the residual (one-step-ahead prediction error) sequence uncorrelatedness (whiteness) hypothesis \cite[pp. 512-513]{Ljung99}.

\subsection{Global modeling under multiple flight states}

The VFP representation allows for complete and precise modeling of the global dynamics under multiple flight states with each state defined --within this study-- by a specific airspeed and AoA that form the flight state vector $\bld{k}$. The VFP model structure allows the functional dependence of model parameters and residual series covariance on both the airspeed and AoA (flight state vector $\bld{k}$). Furthermore, the interrelations and statistical dependencies between the data records corresponding to the different flight states are also taken into account within this structure.  

The VFP-AR representation belongs to the recently introduced broader class of stochastic FP models, which makes use of functional data pooling techniques for combining and optimally treating (as one entity) the data obtained from various experiments corresponding to different structural states, and statistical techniques for model estimation \cite{Kopsaftopoulos12, Kopsaftopoulos-Fassois12, Sakellariou-Fassois16}.

The global modeling via a VFP-AR model involves consideration of all admissible flight states, i.e. airspeeds and AoA, that define the flight envelope. A total of $M_1 \times M_2$ experiments are performed (physically or via coupled fluid-structure interaction computational models and corresponding simulations), with $M_1$ and $M_2$ designating the number of experiments under the various airspeeds and AoA, respectively. Each experiment is characterized by a specific airspeed $k^1$ and a specific AoA $k^2$, with the complete series covering the required range of each variable, say $[k^1_{min}, \: k^1_{max}]$ and $[k^2_{min}, \: k^2_{max}]$, via the discretizations $\{k^1_1, k^1_2,\ldots, k^1_{M_1}\}$ and $\{k^2_1, k^2_2,\ldots, k^2_{M_2}\}$. For the identification of a global VFP model the flight state vector $\bld{k}$ containing the airspeed and AoA components, is formally defined as:

\beq \bld{k} = [k^1_i \; k^2_j]^T \Longleftrightarrow k_{i,j}, \quad i=1,\ldots,M_1, \quad j=1,\ldots,M_2 \eeq 
with $k_{i,j}$ designating the flight state of the wing corresponding to the $i$-th airspeed and the $j$-th AoA. This procedure yields a pool of response signals (each of length $N$):

\beq y_{\bld{k}}[t] \;  \; \mbox{with} \; t=1, \ldots,N, \; k^1 \in \{k^1_1, \ldots, k^1_{M_1}\}, \; k^2 \in \{k^2_1, \ldots, k^2_{M_2} \}. \label{eq:data} \eeq

\begin{figure}[t!]
    \centering
    \includegraphics[scale=0.5]{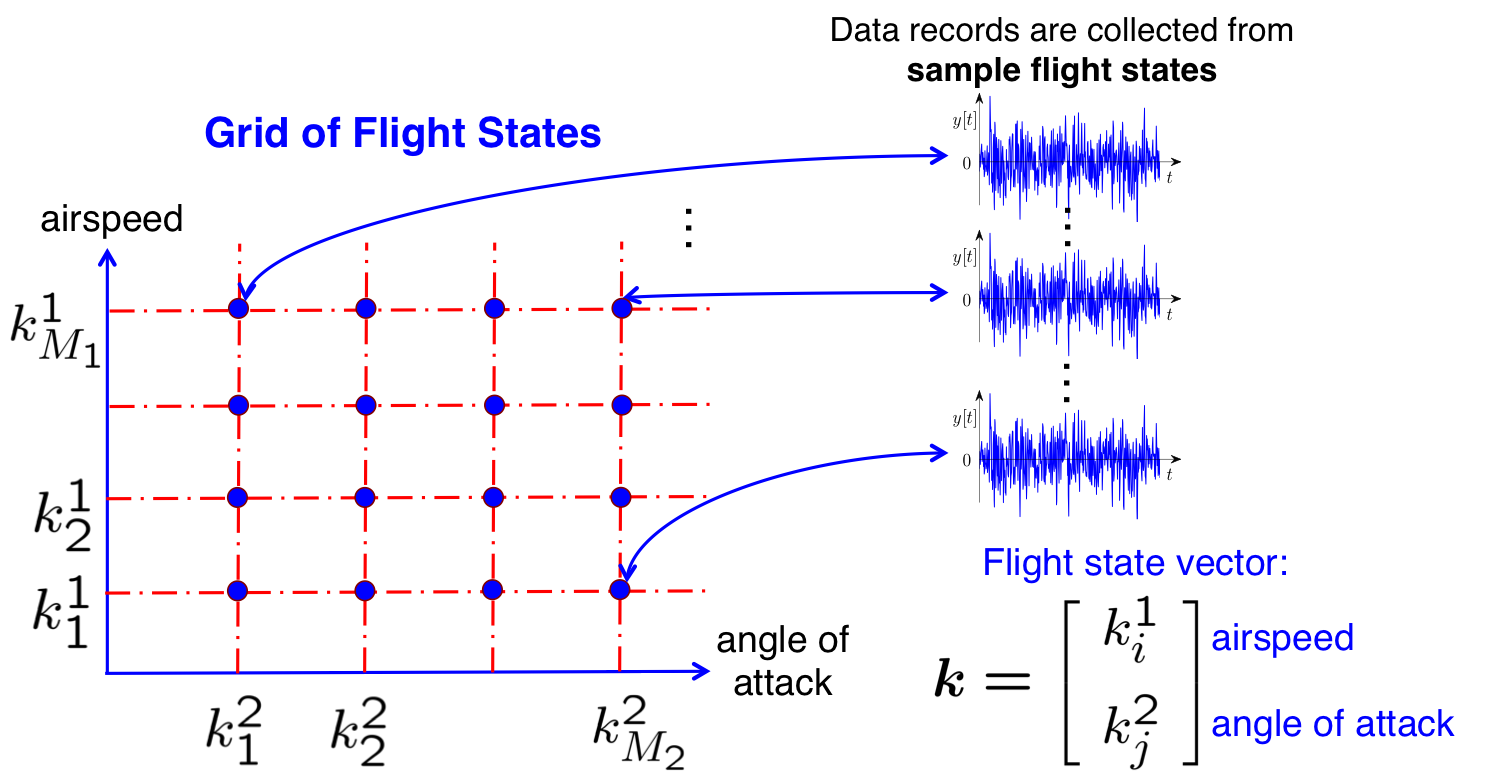}
    \caption{\label{fig:data-grid} \textbf{Grid of flight states for VFP model identification.} Schematic representation of data collection for the identification of a VFP model under different flight states characterized by varying airspeed and AoA.}
\end{figure}

A schematic representation of the data collection process for the identification of the global VFP model is presented in Figure \ref{fig:data-grid}.

A proper mathematical description of the global dynamics under varying flight states may be then obtained in the form of a VFP-AR model. In the case of several response measurement locations an array of such models (or else a vector model) may be obtained, with each scalar model corresponding to each measurement location.

The VFP-AR model is of the following form \cite{Kopsaftopoulos12}:
%
%
\beq y_{\bld{k}}[t] + \sum_{i=1}^{n} a_i({\bld{k}}) \cdot y_{\bld{k}}[t-i] = e_{\bld{k}}[t]  \label{eq:vfp-ar} \eeq

\beq e_{\bld{k}}[t]  \sim \, \mbox{iid} \, \mathcal{N} \bigl( 0,\sigma^2_e(\bld{k}) \bigr) \qquad {\bld{k}} \in \mathbb{R}^2 \eeq

\beq E \{ e_{k_{i,j}}[t] \cdot e_{k_{m,n}}[t-\tau] \} = \gamma_e[k_{i,j},k_{m,n}] \cdot \delta[\tau] \label{eq:resid-cov} \eeq
%
%
with $n$ designating the AR order, $y_{\bld{k}}[t]$ the piezoelectric sensor's response signal, and $e_{\bld{k}}[t]$ the model's residual (one-step-ahead prediction error) sequence, that is a white (serially uncorrelated) zero mean sequence with variance  $\sigma^2_e(\bld{k})$. This may potentially be cross-correlated with its counterparts corresponding to different experiments (different $\bld{k}$'s). The symbol $E\{\cdot\}$ designates statistical expectation, $\delta[\tau]$ the Kronecker delta (equal to unity for $\tau=0$ and equal to zero for $\tau \neq 0$), $\, \mathcal{N}(\cdot,\cdot)$ Gaussian distribution with the indicated mean and variance, and iid stands for identically independently distributed.

The uniqueness of the VFP model structure is that the model parameters $a_i({\bld{k}})$ are modeled as explicit functions of the flight vector $\bld{k}$ (which contains the airspeed and AoA components):
\beq a_i({\bld{k}}) = \sum_{j=1}^{p} a_{i,j} \cdot G_j({\bld{k}}). \label{eq:vfp-ar-coef} \eeq

As equation (\ref{eq:vfp-ar-coef}) indicates, the AR parameters $a_i({\bld{k}})$ are functions of the flight vector $\bld{k}$ by belonging to $p$-dimensional functional subspace spanned by the mutually independent basis functions $G_1({\bld{k}}),G_2({\bld{k}}),\ldots,G_{p}({\bld{k}})$ (\emph{functional basis}). The functional basis consists of polynomials of two variables (bivariate) obtained as tensor products from their corresponding univariate polynomials (Chebyshev, Legendre, Jacobi, and other families \cite{Kopsaftopoulos12, Kopsaftopoulos-Fassois12}). The constants $a_{i,j}$ designate the AR coefficients of projection to be estimated from the measured signals.

Using the backshift operator $\mathcal{B}^i$ $\bigl( \mathcal{B} \cdot x[t] \df x[t-i] \bigr)$ the VFP-AR model may be expressed as follows:
%
\beq A[\mathcal{B},\vk] \cdot y_{\vk}[t] = e_{\vk}[t], \label{eq:tf-repr1} \eeq
with $A[\mathcal{B},\vk]$ designating the AR $\vk$-dependent polynomial operator:
\beq A[{\cal{B}},\bld{k}] \df 1 + \sum_{i=1}^{n} a_i(\bld{k}){\cal{B}}^i. \eeq



\begin{figure}[t!]
    \centering
    \includegraphics[scale=0.45]{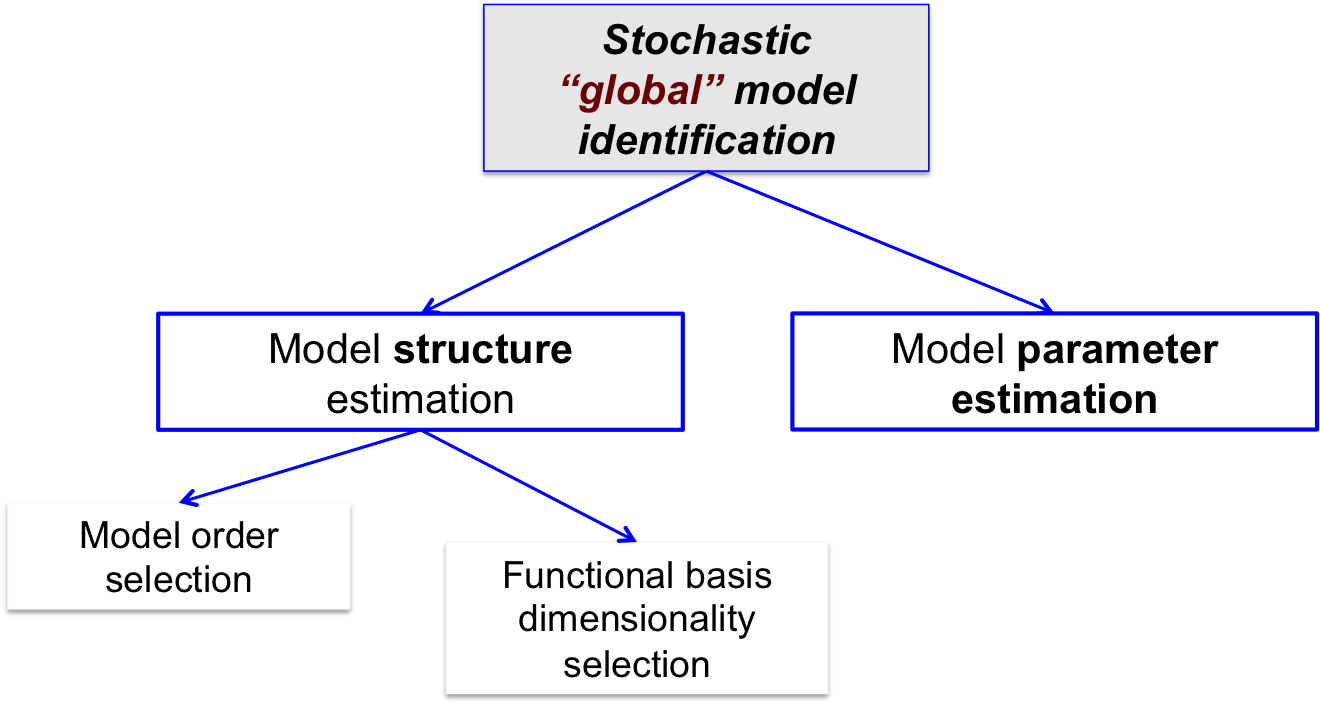}
    \caption{\label{fig:model-id} \textbf{Stochastic VFP identification subtasks.} The VFP model identification is divided into two subtasks: (i) model parameter estimation and (ii) model structure estimation. Model structure estimation addresses the model order and functional basis dimensionality selection.}
\end{figure}

The VFP model identification is divided into two subtasks: (i) model parameter estimation  and (ii) model structure estimation. The VFP identification process is schematically presented in Figure \ref{fig:model-id}.

\subsubsection{Model parameter estimation}

The VFP-AR model of equations (\ref{eq:vfp-ar})--(\ref{eq:vfp-ar-coef}) is parameterized in terms of the parameter vector to be estimated from the measured signals:
\beq {\bld{\bar{\theta}}} = [\; a_{1,1} \; a_{1,2} \; \ldots \; a_{i,j} \; \vdots \; \sigma_e^2(\bld{k}) \;]^T \quad \forall \; \bld{k} \label{eq:param-vec} \eeq
and may be written in linear regression form as:
\beq y_{\bld{k}}[t] = \bigl[ \bld{\varphi}_{\bld{k}}^T[t] \otimes \bld{g}^T(\bld{k}) \bigr] \cdot \bld{\theta} + e_{\bld{k}}[t] = \bld{\phi}_{\bld{k}}^T[t] \cdot \bld{\theta} +  e_{\bld{k}}[t]
\label{eq:lin-reg} \eeq
with:
\begin{subequations}
\bgqar \bld{\varphi}_{\bld{k}}[t] & \df & \Bigl[ -y_{\bld{k}}[t-1] \, \ldots \, -y_{\bld{k}}[t-n] \, \Bigr]^T_{[n\times 1]} \label{eq:varphi} \\
\bld{g}(\bld{k}) & \df & \Bigl[ G_1(\bld{k}) \, \ldots \, G_p(\bld{k}) \Bigr]^T_{[p \times 1]} \\
\bld{\theta} & \df & \Bigl[ a_{1,1} \; a_{1,2} \; \ldots \; a_{n,p} \; \Bigr]^T_{[(np \times 1]} \enqar
\end{subequations}
and $^T$ designating transposition and $\otimes$ Kronecker product \cite[Chap. 7]{Bernstein05}.

Pooling together the expressions of equation (\ref{eq:lin-reg}) of the VFP-AR model corresponding to all flight vectors $\bld{k} \; (k_{1,1}, k_{1,2}, \ldots, k_{M_1,M_2})$ considered in the experiments (cross-sectional pooling) yields:

\beq \left[ \bgary{c}
y_{k_{1,1}}[t] \\ \vdots \\ y_{k_{M_1,M_2}}[t] \enary \right] = \left[ \bgary{c} \bld{\phi}_{k_{1,1}}^T[t] \\
\vdots \\ \bld{\phi}_{k_{M_1,M_2}}^T[t] \enary \right] \cdot \bld{\theta} + \left[ \bgary{c} e_{k_{1,1}}[t] \\
\vdots \\ e_{k_{M_1,M_2}}[t] \enary \right]
\Longrightarrow \bld{y}[t] = \bld{\Phi}[t] \cdot \bld{\theta} + \bld{e}[t] \label{eq:regr-all-k} .\eeq

Then, following substitution of the data for $t=1,\ldots,N$ the following expression is obtained:

\beq \bld{y} = \bld{\Phi} \cdot \bld{\theta} + \bld{e} \label{eq:regr-all-k-t} \eeq
with
\beq \bld{y} \df \left[ \bgary{c} \bld{y}[1]
\\ \vdots \\ \bld{y}[N] \enary \right], \quad
\bld{\Phi} \df \left[ \bgary{c} \bld{\Phi}[1] \\ \vdots \\ \bld{\Phi}[N]
\enary \right], \quad \bld{e} \df \left[ \bgary{c} \bld{e}[1] \\ \vdots \\ \bld{e}[N] \enary \right] \label{eq:regr-all-k-matrix} .\eeq

Using the above linear regression framework the simplest approach for estimating the projection coefficients vector $\bld{\theta}$ is based on minimization of the Ordinary Least Squares (OLS) criterion: 
\beq J^{\OLS} = \frac{1}{N} \sum_{t=1}^{N} \bld{e}^T[t] \bld{e}[t]. \eeq

A more appropriate criterion is (in view of the Gauss-Markov theorem \cite{Greene03}) the Weighted Least Squares (WLS) criterion:
\beq J^{\WLS} = \frac{1}{N} \sum_{t=1}^{N} \bld{e}^T[t]\bld{\Gamma}_{\bld{e}[t]}^{-1}\bld{e}[t] = \frac{1}{N} \bld{e}^T \bld{\Gamma}_{\bld{e}}^{-1} \bld{e} \label{eq:J-wls} \eeq
which leads to the \emph{Weighted Least Squares (WLS)} estimator:
\beq \widehat{\bld{\theta}}^{\WLS} = \bigl[\bld{\Phi}^T \bld{\Gamma}_{\bld{e}}^{-1} \bld{\Phi}\bigr]^{-1} \bigl[\bld{\Phi}^T \bld{\Gamma}_{\bld{e}}^{-1} \bld{y} \bigr] \label{eq:wls-estim}.\eeq
In these expressions $\bld{\Gamma}_{\bld{e}}=E\{\bld{e}\bld{e}^T\}$ ($\bld{\Gamma}_{\bld{e}}=\bld{\Gamma}_{\bld{e}[t]} \otimes \bld{I}_N$, with $\bld{I}_N$ designating the $N \times N$ unity matrix) designates the residual covariance matrix, which is practically unavailable. Nevertheless, it may be consistently estimated by applying (in an initial step) Ordinary Least Squares (details in \cite{Kopsaftopoulos12}). Once $\widehat{\bld{\theta}}^{\WLS}$ has been obtained, the final residual variance and residual covariance matrix estimates are obtained as:
\beq \widehat\sigma_{e}^2(\bld{k}, \widehat{\bld{\theta}}^{\WLS}) = \frac{1}{N} \sum_{t=1}^{N} e^2_{\bld{k}}[t, \widehat{\bld{\theta}}^{\WLS}], \quad \widehat{\bld{\Gamma}}_{\bld{e}[t]} = \frac{1}{N} \sum_{t=1}^{N} \bld{e}[t,\widehat{\bld{\theta}}^{\WLS}] \bld{e}^T[t,\widehat{\bld{\theta}}^{\WLS}] \label{eq:var-cov-est} .\eeq

The estimator $\widehat{\bld{\theta}}^{\WLS}$ may, under mild conditions, be shown to be asymptotically Gaussian distributed with mean coinciding with the true parameter vector $\bld{\theta}^o$ and covariance matrix $\bld{P_{\theta}}$ \cite{Kopsaftopoulos12}:

\beq \sqrt{N}(\widehat{\bld{\theta}}_N - \bld{\theta}^o) \; \sim \; \mathcal{N}(\bld{0},\bld{P_{\theta}}) \quad (N \longrightarrow \infty) \eeq
based on which interval estimates of the true parameter vector may be constructed \cite{Kopsaftopoulos12, Kopsaftopoulos-Fassois12}.

\subsubsection{Model structure estimation}

The problem of VFP-AR model structure estimation (structure selection) for a given family of basis functions (such as Chebyshev, Legendre, and so on) refers to the model order determination for the AR polynomial and the determination of the corresponding functional subspace. Usually, the AR model order is initially selected via customary model order selection techniques (BIC, RSS, frequency stabilization diagrams) \cite{Ljung99}. On the other hand, the functional subspace dimensionality may be selected via a similar BIC-based process in the case of ``complete'' (that is including all consecutive basis functions up to the specified degree) functional subspace \cite{Kopsaftopoulos12, Kopsaftopoulos-Fassois12}, or via the use of a hybrid Genetic Algorithm (GA) procedure in the case of ``incomplete'' (that is not necessarily including all consecutive basis functions up to the specified degree) functional subspace \cite{Kopsaftopoulos12, Kopsaftopoulos-Fassois12}. For the latter approach, initially the maximum functional subspace dimensionality is selected, which defines the search space of the functional subspace estimation subproblem. The determination of the exact subspace dimensionality is achieved via the use of GAs based on minimization of the BIC with respect to the candidate basis functions. In the current study, the estimation of the functional subspace dimensionality is achieved via the use of the BIC criterion for increasing functional subspace dimensionality. The functional basis consists of bivariate Chebyshev Type II polynomials \cite{Dunkl-Xu01, Kowalski82, Krall-Scheffer67}. 


\section{The Bio-inspired Self-sensing Composite Wing} \label{sec:wing-sensors} 

The prototype bio-inspired self-sensing composite wing was designed and fabricated at Stanford University. It is outfitted with micro-fabricated multi-modal distributed sensor networks that have been embedded between the carbon-fiber and fiberglass layers of the top composite skin of the wing structure. The composite wing with the embedded micro-sensor networks constitutes a self-sensing structure that with the integration of appropriate algorithms and corresponding software is able to achieve high-resolution state and structural awareness along with self-diagnostic capabilities. 

\subsection{Bio-inspired stretchable sensor networks} \label{sec:sensors} 

\begin{figure}[t!]
    \centering
    \includegraphics[scale=1.3]{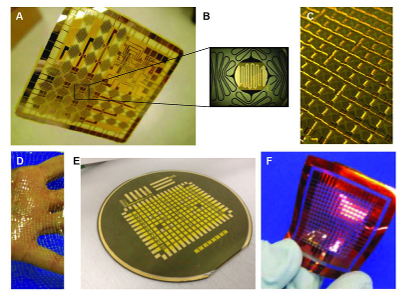}
    \caption{\label{fig:sensors} \textbf{Bio-inspired stretchable sensor networks.} (A) A 16-node sensor network on a wafer can be expanded up to $1,057 \%$ in each dimension after release. (B) Close-up of the sensor node demonstrating the design of the microwires. (C) A sensor network with 169 nodes before expansion. (D) An expanded 5041-node network is shown in contrast to a hand, which illustrates the flexibility of the membrane. (E) Network before release on a 100 mm wafer. (F) A fabricated 256-node network on polyimide is easily held by hand without damaging the network. It is characterized by 16 $\mu m$ wide, 50 $\mu m$ thick microwires.}
\end{figure}

The bio-inspired stretchable sensor networks used in this study are developed via the use of nonstandard micro/nano-fabrication CMOS (complementary metal-oxide-semiconductor) and MEMS (micro-electro-mechanical) processes. They consist of various sensor types (piezoelectric \cite{Salowitz-etal13, Salowitz-etal13b}, strain \cite{Salowitz-etal13, Guo14}, temperature \cite{Lanzara-etal10, Salowitz-etal13, Guo-etal11, Guo14}, and pressure sensors), i.e. multi-modal sensor networks, and can be installed monolithically into host materials, either embedded between the layers of composite materials or mounted on the structure of metallic or composite structures. Recently, micro-fabricated stretchable sensor networks have been developed and deployed micro-scale sensors over macroscopic areas \cite{Lanzara-etal10, Salowitz-etal14, Salowitz-etal13, Salowitz-etal-IWSHM13, Guo-etal11, Guo14, Salowitz-etal13b}. In order to survive the large deformation and strains that occur during the stretching process, the sensors are created on polymer substrates with nonstandard and unique micro/nano-fabrication processes \cite{Lanzara-etal10, Salowitz-etal13, Guo14}. The resulting sensors and wiring components have dimensions in the order of tens of micrometers. Figure \ref{fig:sensors} presents some indicative bio-inspired network designs fabricated at Stanford University. The network used in this study corresponds to the 256-node design shown in Figure \ref{fig:sensors}E.

The network is created on standard 100 mm diameter substrates and expanded to span areas orders of magnitude larger than the initial fabrication area, thus enabling the deployment of hundreds of micro-meter scale devices over meter-scale areas. The resulting web-like network --mimicking biological nervous systems-- consists of distributed small scale components (sensor nodes, communication nodes, wires, and connection pads) intended to have a minimal parasitic effect on the host structure without causing any negative impact on the mechanical properties and dynamic behavior. The component size is on the same order as an individual fiber in typical composite materials and small enough to be embedded inside composites without any additional structural modifications. These networks can be used in-situ, from the material fabrication throughout its service life, to monitor the curing process of composite materials, characterize material properties post-cure, and monitor the structural dynamics, performance, and health of the structure.

In this work, four stretchable multi-modal sensor networks consisting of distributed piezoelectric lead zirconate titanate (PZT) sensors, strain gauges, and resistive temperature detectors (RTD) sensors have been designed and fabricated \cite{Lanzara-etal10, Salowitz-etal14, Salowitz-etal13, Guo-etal11, Guo14, Salowitz-etal13b} so that they can be embedded inside the composite layup of the top skin of the wing. Each of the four sensor networks contains 8 piezoelectric sensors (disc PZT $3.175$ mm in diameter), 6 strain gauges, and 24 RTDs. The total number of embedded sensors in the composite wing is 148. Stretchable wires connect the network nodes and serve as signal communication channels. Before stretching, the network dimensions are $52.8$ mm by $39.6$ mm, while after the stretching process expands to 140 mm by 105 mm yielding a $700 \%$ total surface area increase \cite{Guo14}. After the stretching process takes place, the outer network pads are connected to a surrounding flexible PCB that facilitates the connection with the data acquisition system.

\subsection{The self-sensing composite wing} \label{sec:wing}

The prototype wing was designed, constructed and tested at Stanford University. Analyzing a prototype model with construction typical of that of an operational UAV wing allows the comparison of the aeroelastic behavior, structural dynamics, and performance characteristics with existing systems, as well as enables a scaling analysis. The wing design is based on the cambered SG6043 high lift-to-drag ratio airfoil with a $0.86$ m half-wingspan, $0.235$ m chord, and an aspect ratio of $7.32$. Table \ref{tab:wing} presents the wing geometry and dimensions. In order to achieve the successful fabrication of the wing prototype, an appropriate network-material integration process had to be developed for embedding the micro-fabricated sensor networks inside the composite materials. 

\begin{table}[b!]
\caption{Self-sensing composite wing geometry. \label{tab:wing}} 
\centering {\small     
\begin{tabular}{ll} 
\hline
\rule[-1ex]{0pt}{3.5ex}  Semispan $b$ & 0.86 m \\
\rule[-1ex]{0pt}{3.5ex}  Chord $c$ & 0.235 m \\
\rule[-1ex]{0pt}{3.5ex}  Area $S$ & 0.2 m$^2$    \\
\rule[-1ex]{0pt}{3.5ex}  Aspect Ratio & 7.32  \\
\rule[-1ex]{0pt}{3.5ex}  Airfoil & SG6043  \\
\hline 
\end{tabular} }
\end{table}

\begin{figure}[t!]
    \centering
    \includegraphics[scale=0.9]{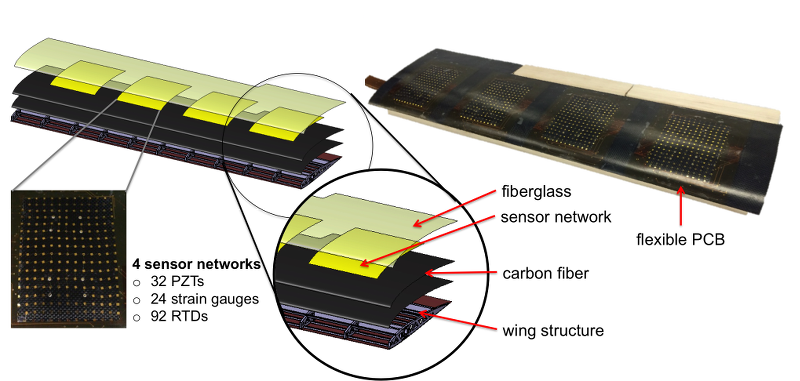}
    \caption{\label{fig:wing} \textbf{Self-sensing intelligent composite wing.} The composite wing design, layers, flexible PCBs, and four networks with a total of 148 (32 piezoelectric, 24 strain gauges, and 92 RTDs) micro-sensors embedded in the composite layup.}
\end{figure}

In order to tackle the wing-network integration challenges related to the micro-scale and fragile nature of the network nodes and wires, a new multi-stage process had to be developed for the transfer, electrical interfacing and insulation of the network components based on multilayer flexible PCBs and epoxy armoring. The composite wing structure was manufactured from carbon fiber and fiberglass laminated composites. The wing skin layup consists of carbon fiber (CF) plain wave fabric 1K T300 and fiberglass (FG) plain wave fabric 18 gr/m$^2$ infused with Araldite LY/HY5052 epoxy. The stacking sequence of the layers is [0$^o$ FG, 0$^o$ CF, 45$^o$ CF, 45$^o$ CF, 0$^o$ CF, 0$^o$ FG]. The wing design, layers, flexible PCBs, and sensor networks are shown in Figure \ref{fig:wing}). The four networks are embedded between the two top layers at $0^o$ of the layup (near the wing surface) during the lamination process. The fiberglass was employed due to its transparency, so that the embedded stretchable sensor networks are visible to the naked eye. The supporting wing structure consists of basswood spars and ribs upon which the composite skin is adhesively mounted.

\section{The Wind Tunnel and the Experiments} \label{sec:experiments}

\subsection{The wind tunnel}

\begin{figure}[t!]
    \centering
    \includegraphics[scale=1.5]{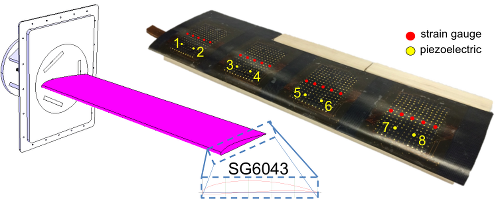}
    \caption{\label{fig:wing-sensors} \textbf{Wing and sensor locations.} The wing airfoil and the locations of 8 piezoelectric and 20 strain sensors.}
\end{figure}

The prototype composite wing was tested in the open-loop wind tunnel facility at Stanford University. The wind tunnel has a square test section of 0.84 $\times$ 0.84 m (33 $\times$ 33 in) and can achieve continuous flow speeds up to approximately 40 m/s. A custom basis was designed and fabricated to support the wing and permit adjustments in the AoA. The wing was mounted horizontally inside the test section. The axis of rotation coincides with the quarter of the wing chord. Figure \ref{fig:wing-sensors} presents the locations of 8 piezoelectric and 20 strain sensors on the composite wing. 

￼
\subsection{Experiments under varying flight states}

A series of wind tunnel experiments were conducted for various angles of attack and freestream velocities $U_{\infty}$. For each AoA, spanning the range from 0 degrees up to 18 degrees with an incremental step of 1 degree, data were sequentially collected for all velocities within the range 9 m/s to 22 m/s (incremental step of 1 m/s). The above procedure resulted in a grid of flight state data sets corresponding to 266 different experiments covering the complete range of the considered flight states. The experimental flight states along with the corresponding Reynolds numbers are outlined in Table \ref{tab:experiments}.

\subsection{The signals}

For each experiment the vibration response was recorded at different locations on the wing via the embedded piezoelectric sensors (initial sampling frequency $f_s = 1000$ Hz, initial signal bandwidth $0.1 - 500$ Hz). The signals were recorded via a National Instruments X Series 6366 data acquisition module featuring eight 16-bit simultaneously sampled analog-to-digital channels. After a preliminary investigation, the response signal bandwidth for the parametric analysis is selected as $0.1 - 80$ Hz. The initial signals are low-pass filtered (Chebyshev Type II) and sub-sampled to a resulting sampling frequency $f_s = 200$ Hz. Table \ref{tab:signal-details} summarizes the piezoelectric data acquisition, signal, and pre-processing details.

\begin{table}[t!]
\caption{The flight states considered in the wind tunnel experiments. For each constant AoA within the range of $[0 - 18]$ degrees, a series of data sets was recorded corresponding to freestream velocities $[9 - 22]$ m/s. \label{tab:experiments}} 
\begin{center}       {\small
\begin{tabular}{lcccccccccccccc} 
\hline
\rule[-1ex]{0pt}{3.5ex}  $Re \ (\times 10^3)$ & 124 & 155 & 171 & 187 & 202 & 217 & 233 & 248 & 264 & 280 & 295 & 311 & 326 & 342 \\
\hline
\rule[-1ex]{0pt}{3.5ex}  $U_{\infty}$ (m/s) &  9 & 10 & 11 & 12 & 13 & 14 & 15 & 16 & 17 & 18 & 19 & 20 & 21 & 22 \\
\hline
\rule[-1ex]{0pt}{3.5ex}  & \multicolumn{14}{r}{AoA: 0 -- 18 degrees; Total number of experiments: 266}  \\
\end{tabular} }
\end{center} 
\end{table}

\begin{table}[b!]
\caption{Piezoelectric data acquisition, signal, and pre-processing details.} \label{tab:signal-details} 
\centering {\small
\begin{tabular}{ll}
\hline
Number of sensors: & 8  \\
Sampling frequency: & $f_s = 1000$ Hz  \\ 
Signal length: & $N = 90,000$ samples ($90$ s)  \\
Initial Bandwidth: & $[0.1 - 500]$ Hz  \\
Filtering: & Low-pass Chebyshev Type II (12th order; cut-off frequency 80 Hz)  \\
Filtered Bandwidth: & $[0.1 - 80]$ Hz  \\
\hline
\end{tabular} }
\end{table}

\section{Results and Discussion} \label{sec:results}

\subsection{Numerical simulations}

In order to investigate the theoretical aerodynamic behavior of the fabricated composite wing based on which the experimental results could be interpreted and assessed, a series of numerical simulations was conducted using XFOIL, an interactive software developed at MIT for the design and analysis of subsonic isolated airfoils \cite{XFOIL}. 

\begin{figure}[t]
    \centering
    \includegraphics[scale=0.7]{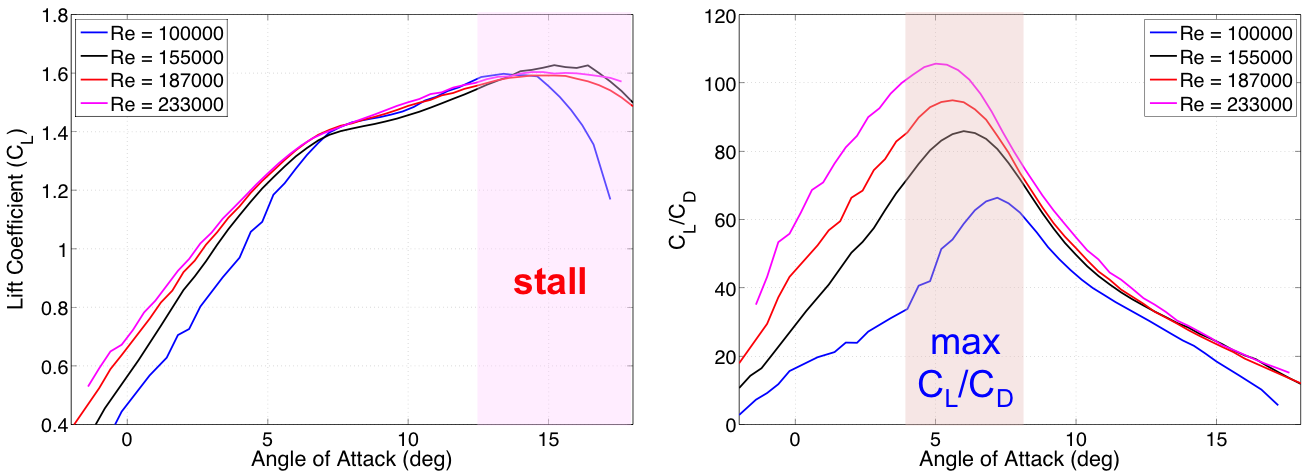}
    \caption{\label{fig:xfoil} \textbf{CFD simulation results.} Indicative simulation results: (a) lift coefficient $C_L$ (left subplot) and (b) lift-to-drag coefficient ratio $C_L/C_D$ (right subplot) versus AoA for the SG6043 airfoil and various Reynolds numbers.}
\end{figure}

Figures \ref{fig:xfoil}a and Figure \ref{fig:xfoil}b present the lift coefficient versus the AoA and lift to drag coefficient ratio $C_L/C_D$ results of the SG6043 airfoil, respectively, for various Reynolds numbers ($U_{\infty} =$ 7, 10, 12 and 15 m/s; see Table \ref{tab:experiments}). It may be readily observed that the wing exhibits stall (loss of lift shown as shaded area in Figure \ref{fig:xfoil}a) starting from an AoA of approximately 12 degrees for a Reynolds number of $Re = 100,000$. As the Reynolds number increases (for increasing freestream velocity) stall occurs for an increasingly higher AoA up to a value of 16 degrees. Moreover, observe that the maximum $C_L/C_D$ ratio is obtained for angles between 4 and 8 degrees (shaded areas in Figure \ref{fig:xfoil}b).

\begin{figure}[t!]
    \centering
    \hc{-0.8}\includegraphics[scale=0.5]{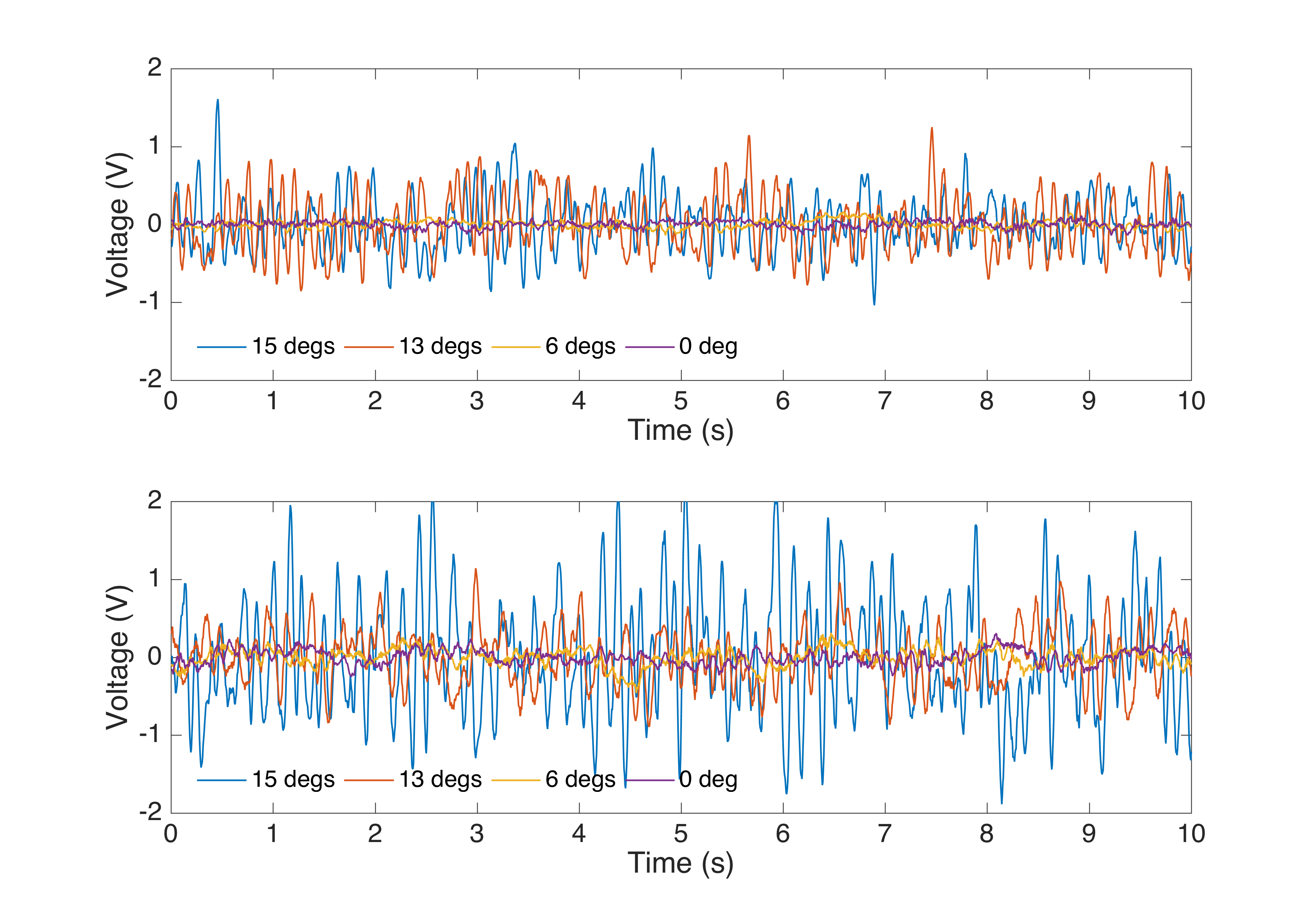}
    \caption{\label{fig:piezo-signals} \textbf{Piezoelectric signals.} Indicative signals obtained from piezoelectric sensor 2 under various angles of attack: (a) freestream velocity $U_{\infty} = 11$ m/s (top subplot) and (b) freestream velocity $U_{\infty} = 15$ m/s (bottom subplot).}
\end{figure}

\subsection{Signals and statistical energy analysis}

Figure \ref{fig:piezo-signals} presents indicative wind tunnel signals obtained from piezoelectric sensor 2 (see Figure \ref{fig:wing-sensors} for the sensor location) under various angles of attack and freestream velocities of $U_{\infty} = 11$ m/s (top subplot) and $U_{\infty} = 15$ m/s (bottom subplot). Observe the stochastic (random) nature of these signals, which is due to the wind tunnel airflow actuation and the aeroelastic response of the wing. In addition, it is evident that for higher angles of attack and as the wing approaches stall, the signal amplitude (voltage) increases. In the case of $U_{\infty} = 11$ m/s (top subplot) in Figure \ref{fig:piezo-signals}, the maximum signal amplitude for AoA of 13 and 15 degrees seems to be similar as there is no evident further increase. For this freestream velocity and based on Figure \ref{fig:xfoil}, stall occurs at an AoA of approximately 13 degrees. In the case of  $U_{\infty} = 15$ m/s (bottom subplot) in Figure \ref{fig:piezo-signals}, stall occurs at approximately 15 degrees, and it may be readily observed that there is an obvious increase in the signal amplitude from 13 to 15 degrees AoA.

\begin{figure}[t!]
    \centering 
    \hc{-0.5}\includegraphics[scale=0.32]{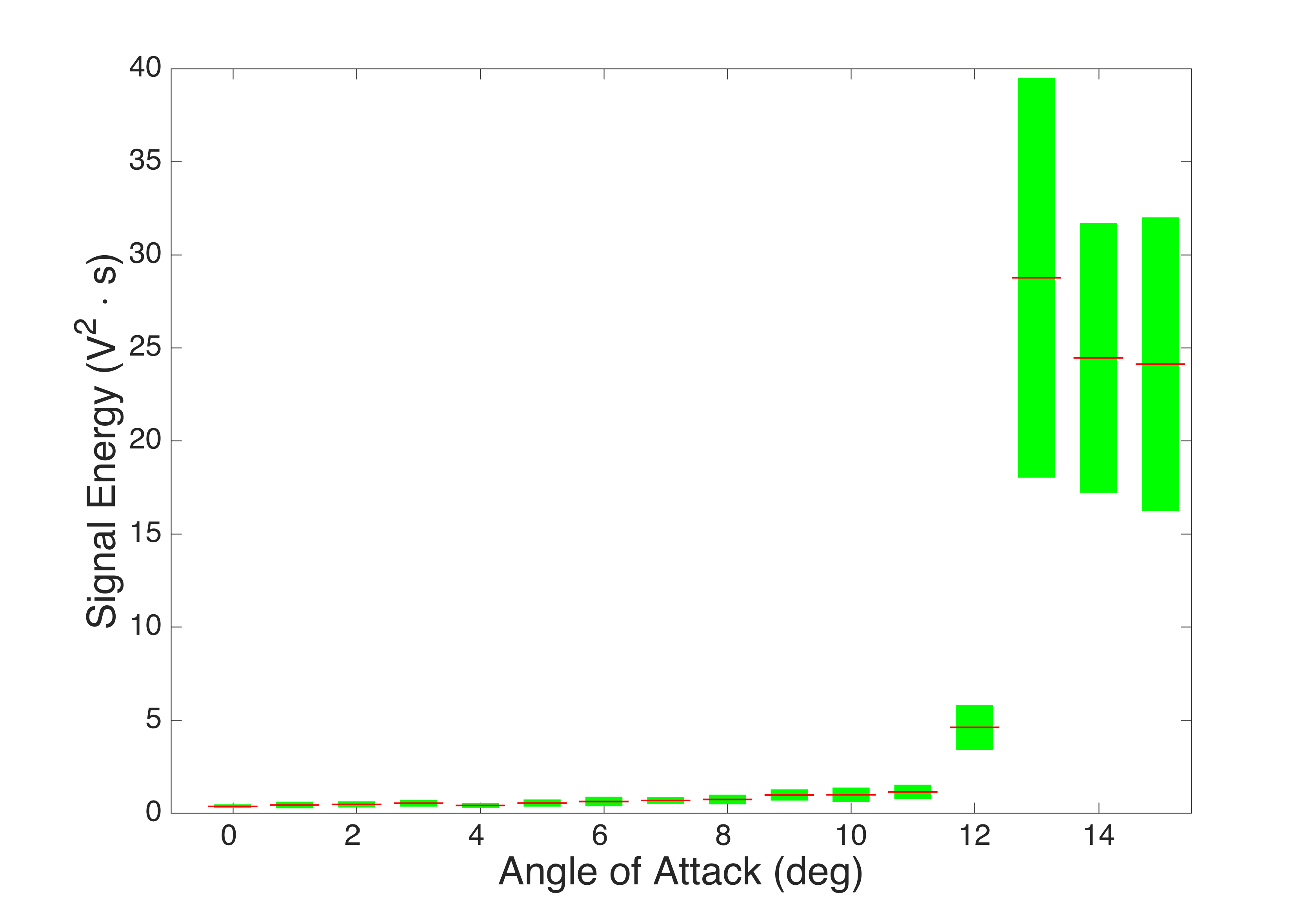}\hc{-0.6}\includegraphics[scale=0.32]{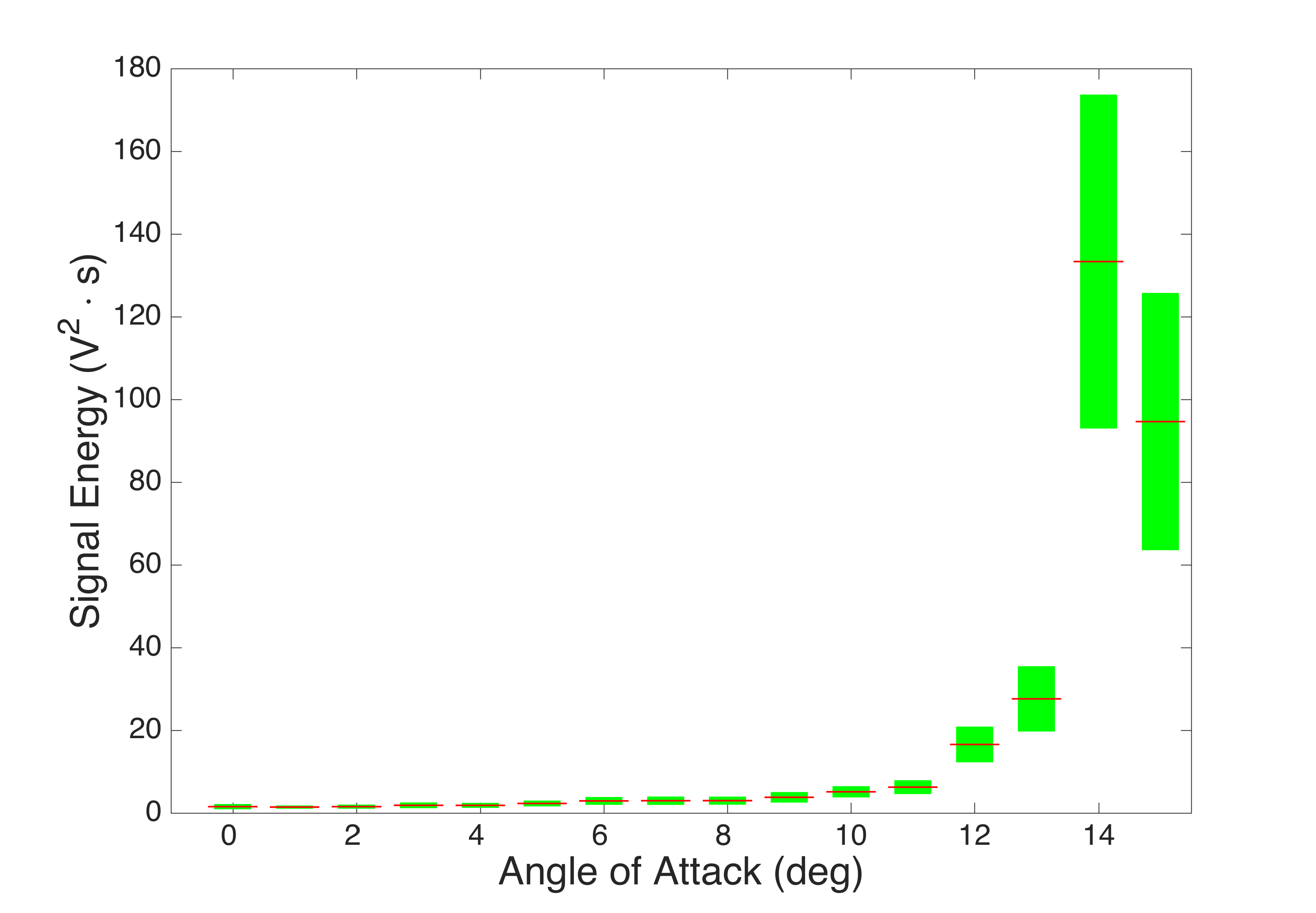}  
    \caption{\label{fig:signal-energy} \textbf{Signal energy statistical analysis for increasing AoA.} Indicative signal energy versus AoA wind-tunnel results for piezoelectric sensor 1 and freestream velocities $U_{\infty} = 11$ m/s (left) and $U_{\infty} = 15$ m/s (right). The mean values of the signal energy are shown as red lines. The $99 \%$ confidence bounds are shown as green shaded areas.}
\end{figure}

In order to further investigate the signal amplitude of the sensors with respect to varying AoA and airspeed we conducted the statistical signal energy analysis based on the wind tunnel experiments. The initial signal of 90 s ($N = 90,000$ samples) was split into  signal windows of 0.5 s ($N = 500$ samples) each. Then, for each signal window the mean value and the standard deviation of the signal energy (time integration of the squared signal $V^2$ within the time window) were estimated. Figure \ref{fig:signal-energy} presents indicative signal energy results obtained from piezoelectric sensor 1 during the wind tunnel experiments. The AoA is varied between 0 and 15 degrees for constant freestream velocities of $U_{\infty} = 11$ m/s (left subplot) and $U_{\infty} = 15$ m/s (right subplot). The goal is to correlate the signal energy in the time domain with the airflow characteristics and aeroelastic properties in order to identify and track appropriate signal features that can be used for the wing vibration monitoring, the localization of the flow separation over the wing chord, and the early detection of stall under various flight states. Figure \ref{fig:signal-energy} presents the mean value of the vibrational signal energy along with the $99 \%$ confidence bounds. 


For the case $U_{\infty} = 11$ m/s (left subplot in \ref{fig:signal-energy}) as the AoA exceeds the value of 12 degrees the signal energy significantly increases and reaches the maximum value as it approaches stall (AoA of 13 degrees). Then, it slightly decreases after stall has occurred (14 and 15 degrees). The sudden increase in the signal energy is caused by the stall-induced oscillations (or stall flutter phenomenon). The statistical analysis of the wind tunnel signals for the various sensors indicated that for velocities in the range of 10 m/s to 12 m/s the stall angle lies within 12 to 13 degrees, whereas for higher velocities the stall AoA may exceed the 15 degrees. The right subplot of Figure \ref{fig:signal-energy} presents similar statistical energy results for freestream velocity $U_{\infty} = 15$ m/s. These results are in agreement with the trend of signals in Figure \ref{fig:piezo-signals} as in both cases the signal amplitude/energy is maximized within the stall range of the wing. Also, the results are in agreement with the numerical simulations presented in Figure \ref{fig:xfoil}.

\begin{figure}[t!]
    \centering 
    \hc{-0.5}\includegraphics[scale=0.32]{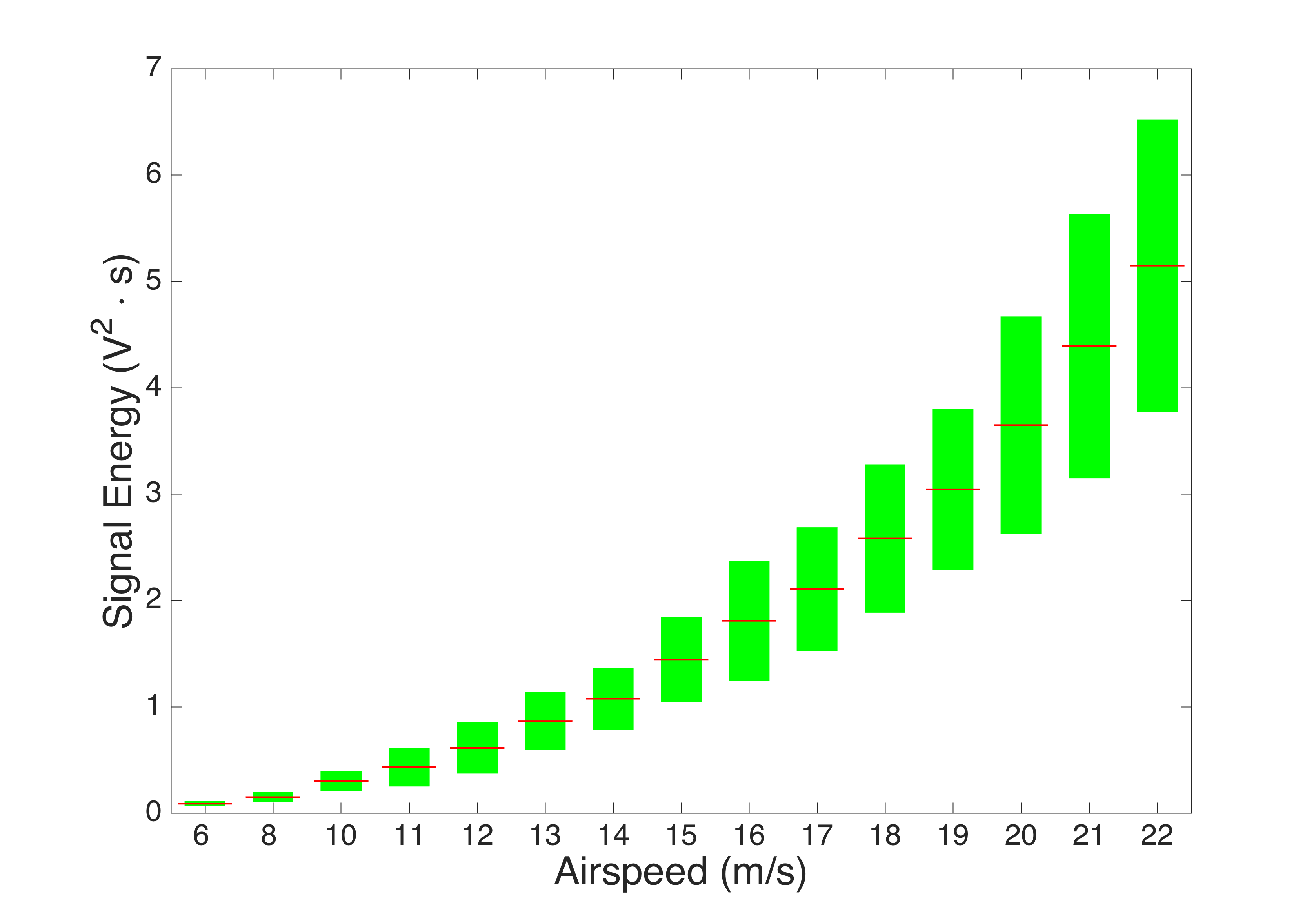}\hc{-0.6}\includegraphics[scale=0.32]{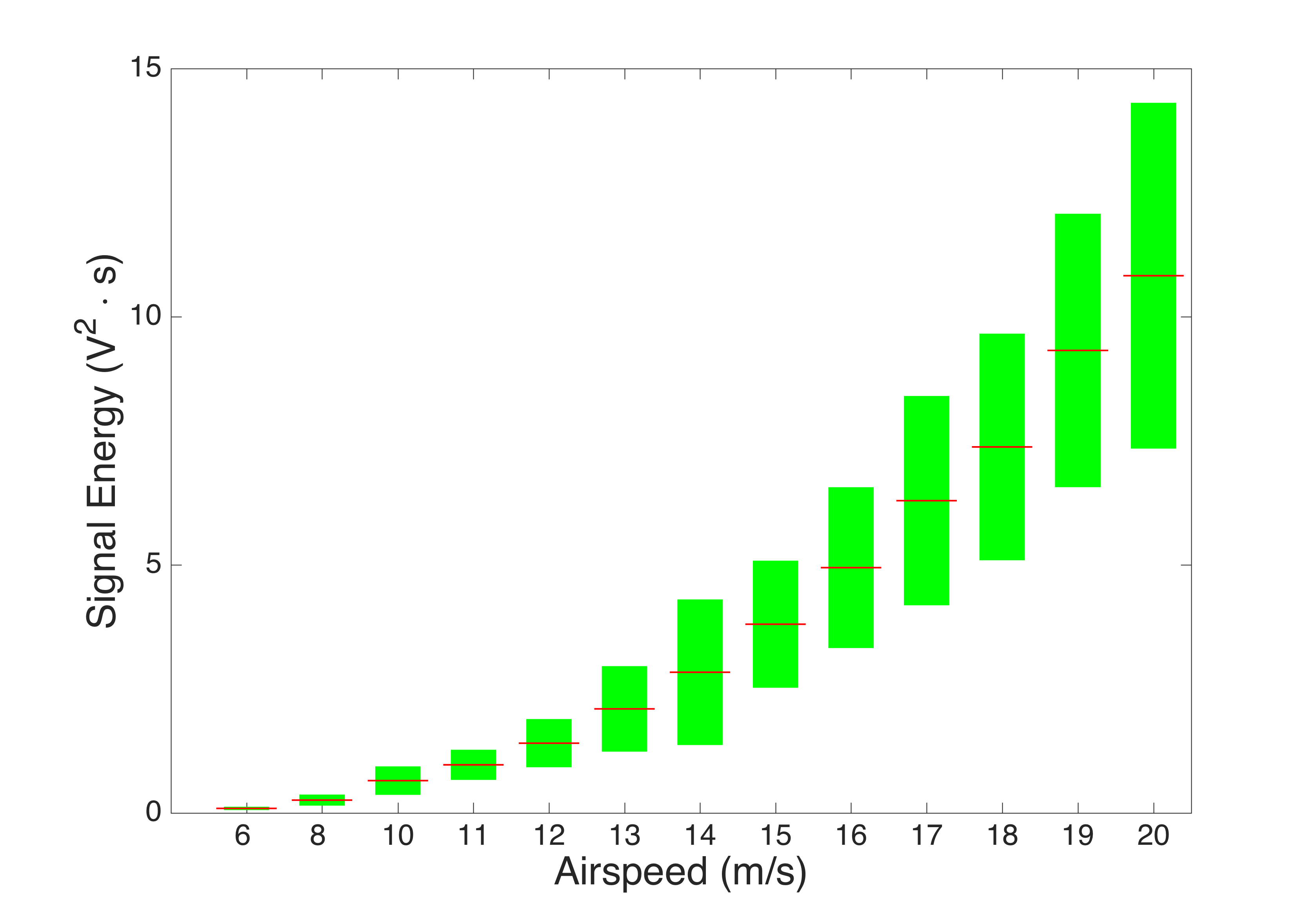}  
    \caption{\label{fig:signal-energy-AS} \textbf{Signal energy statistical analysis for increasing airspeed.} Indicative signal energy versus airspeed wind-tunnel results for piezoelectric sensor 1 and an AoA of 1 degree (left) and 9 degrees (right). The mean values of the signal energy are shown as red lines. The $99 \%$ confidence bounds are shown as green shaded areas.}
\end{figure}

Indicative signal energy statistical analysis results for increasing airspeed are presented in Figure \ref{fig:signal-energy-AS}. The left subplot corresponds to an AoA of 1 degree, while the right subplot to an AoA of 9 degrees. Observe the quadratic increase in the signal energy with respect to increasing airspeed, which is in agreement with the basic aerodynamic lift formula that implies quadratic lift increase for increasing airspeed at a constant AoA. Moreover, observe that as the airspeed increases the confidence bounds also increase. Finally, in agreement with the analysis of Figure \ref{fig:signal-energy}, the signal energy is significantly higher, in fact approximately double, in the case of the 9-degree AoA (right) when compared with the corresponding of 1 degree (right subplot).

\subsection{Non-parametric analysis}

Non-parametric identification is based on $90,000$ ($90$ s) sample-long response signals obtained from the embedded piezoelectric sensors (see Table \ref{tab:signal-details}. A $5096$ sample-long Hamming data window (frequency resolution $\Delta f = 0.24$ Hz) with $90 \%$ overlap is used for the Welch-based spectral estimation (MATLAB function \textit{pwelch.m}). 

\begin{figure}[t!]
    \hc{-0.5}\includegraphics[scale=0.7]{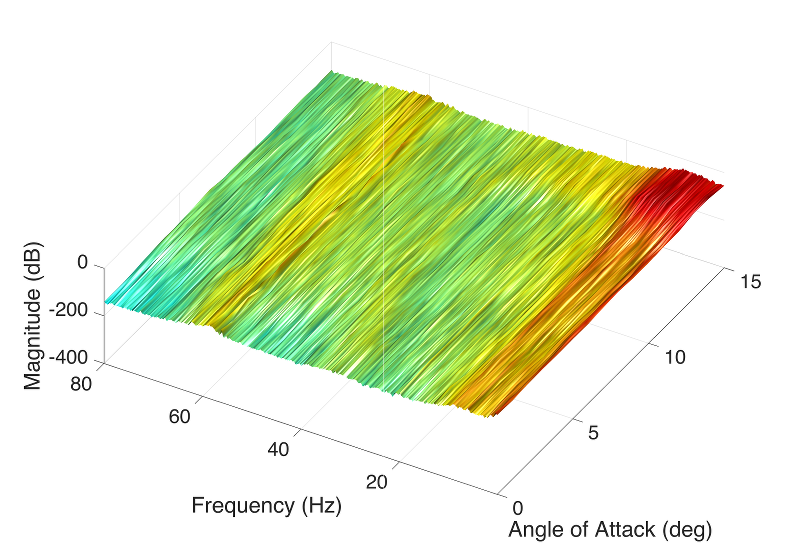}\hc{-0.4}\includegraphics[scale=1.3]{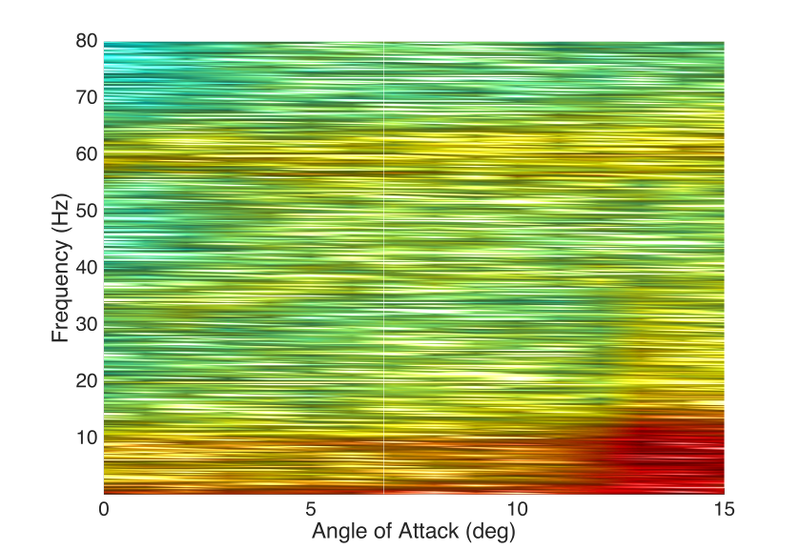}    
    \caption{\label{fig:PSDs-AoA} \textbf{Non-parametric spectral estimates vs AoA.} Indicative non-parametric Welch-based PSD estimates (piezoelectric sensor 1) versus AoA for $U_{\infty} = 13$ m/s ($Re = 202,000$) freestream velocity.}
\end{figure}

Figure \ref{fig:PSDs-AoA} presents indicative non-parametric power spectral density (PSD) Welch-based estimates of the piezoelectric response signals obtained from  sensor 1 for increasing AoA and freestream velocity  $U_{\infty} = 13$ m/s ($Re = 202,000$). Notice that as the AoA increases the PSD amplitude in the lower frequency range of $[0.1 - 12]$ Hz significantly increases as well. More specifically, as the AoA approaches the critical stall range of $[13 - 15]$ degrees, the low frequency vibrations become dominant and thus indicating the proximity to the stall of the wing. From this Figure it is evident that by monitoring the identified lower frequency bandwidths that are sensitive to increasing AoA we may have a strong indication of stall. All the embedded piezoelectric sensors of the wing exhibit a similar performance, but for the sake of brevity the results are presently omitted. 

\begin{figure}[t!]
    \hc{-0.3}\includegraphics[scale=1.3]{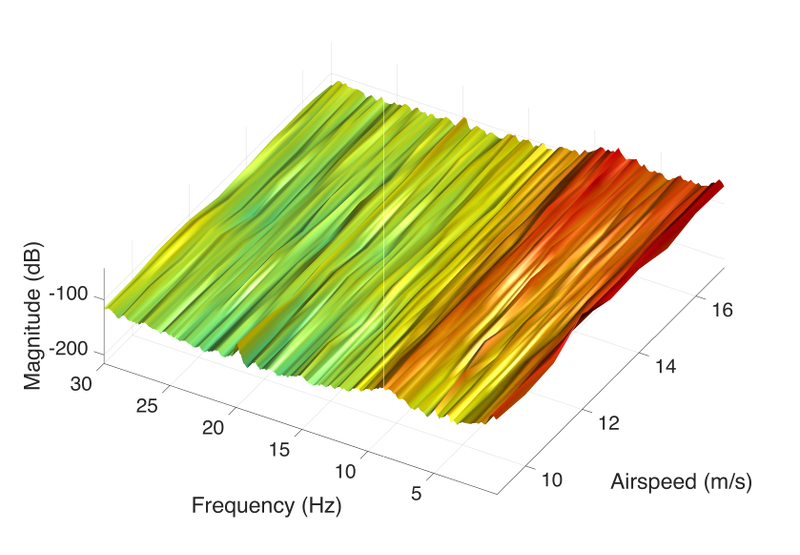}\hc{-0.2}\includegraphics[scale=1.3]{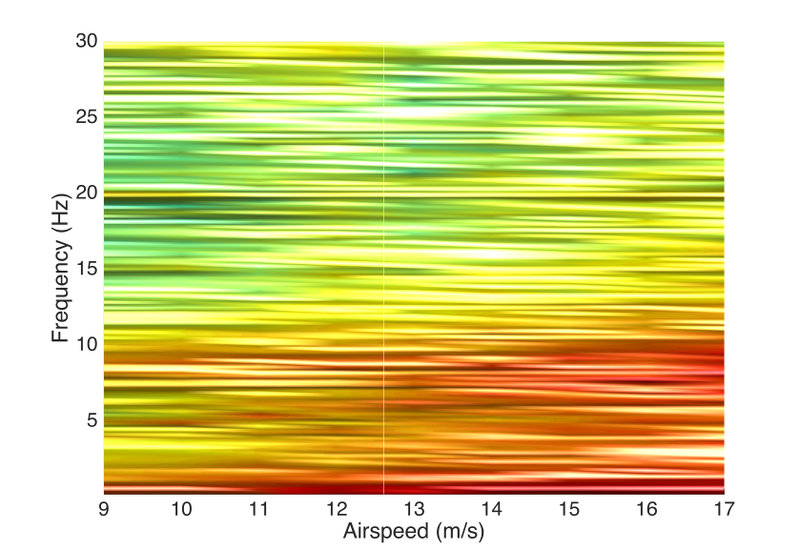}       
    \caption{\label{fig:PSDs-AS} \textbf{Non-parametric spectral estimates vs airspeed.} Indicative non-parametric Welch-based PSD estimates (piezoelectric sensor 1) versus airspeed for an AoA of $0$ degrees.}
\end{figure}

Similarly, Figure \ref{fig:PSDs-AS} presents indicative non-parametric power spectral density (PSD) Welch-based estimates obtained from piezoelectric sensor 1 for increasing airspeed and an constant AoA of 0 degrees within the $[0.5 - 30]$ Hz frequency range. Again, notice that as the airspeed increases, the PSD amplitude in the lower frequency range slightly increases as well. In this case, it is expected that as the airspeed increases for a constant AoA the wing will approach flutter which will be triggered by the coupling of aeroelastic modes. In this case the coupling occurs in $[0.5 - 15]$ Hz frequency range. By carefully observing Figure \ref{fig:PSDs-AS}  it may be seen that the frequency at approximately 5 Hz increases with increasing airspeed and approaches the frequency at approximately 9 Hz, thus providing an indication of incipient flutter. This observation will be clarified by the global parametric modeling results of Section \ref{sec:VFP-results}.

\subsection{Baseline parametric identification} \label{sec:param-analysis}

Conventional AR time series models representing the wing dynamics are obtained through standard identification procedures \cite{Ljung99, Soderstrom-Stoica89} based on the collected piezoelectric response signals (MATLAB function \textit{arx.m}). The response signal bandwidth is selected as $0.1-80$ Hz after the initial signals were low-pass filtered (Chebyshev Type II) and sub-sampled to a resulting sampling frequency $f_s = 200$ Hz (initial sampling frequency was 1000 Hz). Each signal resulted in a length of $N = 4,000$ samples ($20$ s) and was subsequently sample mean corrected (Table \ref{tab:param-signals}). Indicative baseline parametric modeling results are presented for piezoelectric sensor 1 and for a flight state corresponding to an airspeed of 11 m/s and an AoA of 3 degrees.

\begin{figure}[t!]
    \centering
    \includegraphics[scale=0.15]{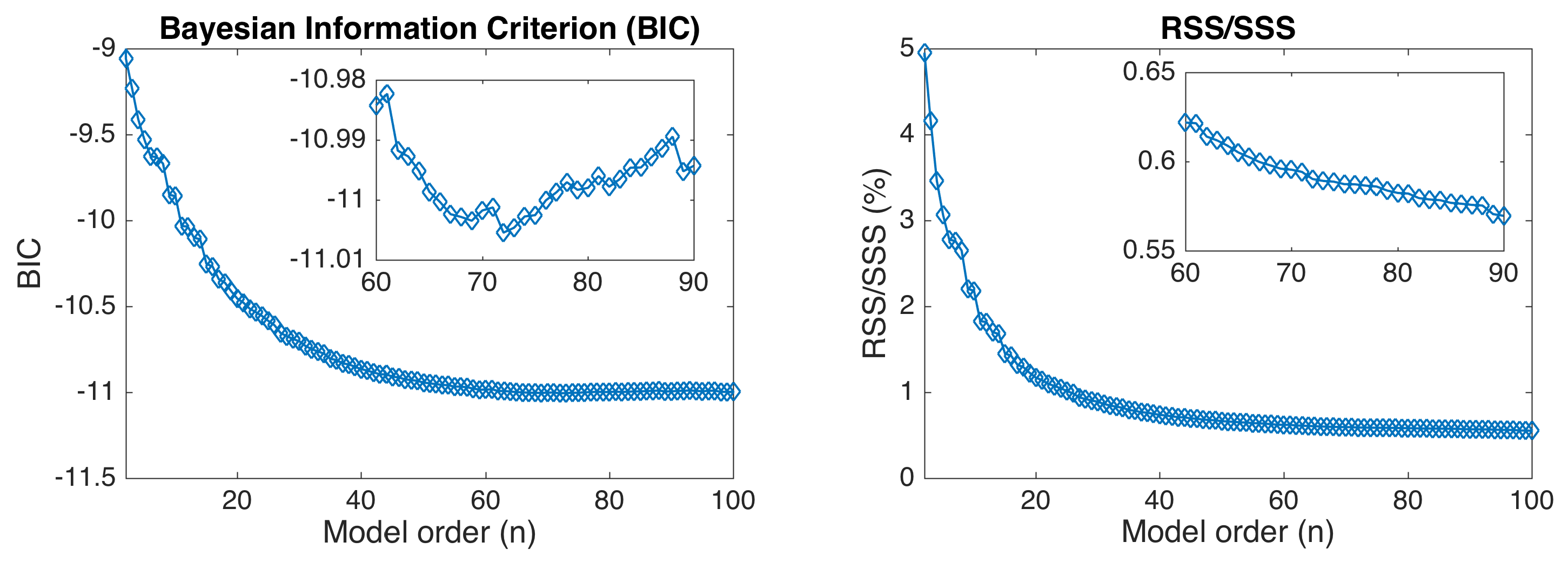}    
    \caption{\label{fig:BIC-RSS} \textbf{AR order selection criteria.} Order selection criteria for AR$(n)$ type parametric models: (a) BIC and (b) RSS/SSS.}
\end{figure}

\begin{figure}[t!]
    \centering
    \includegraphics[scale=0.6]{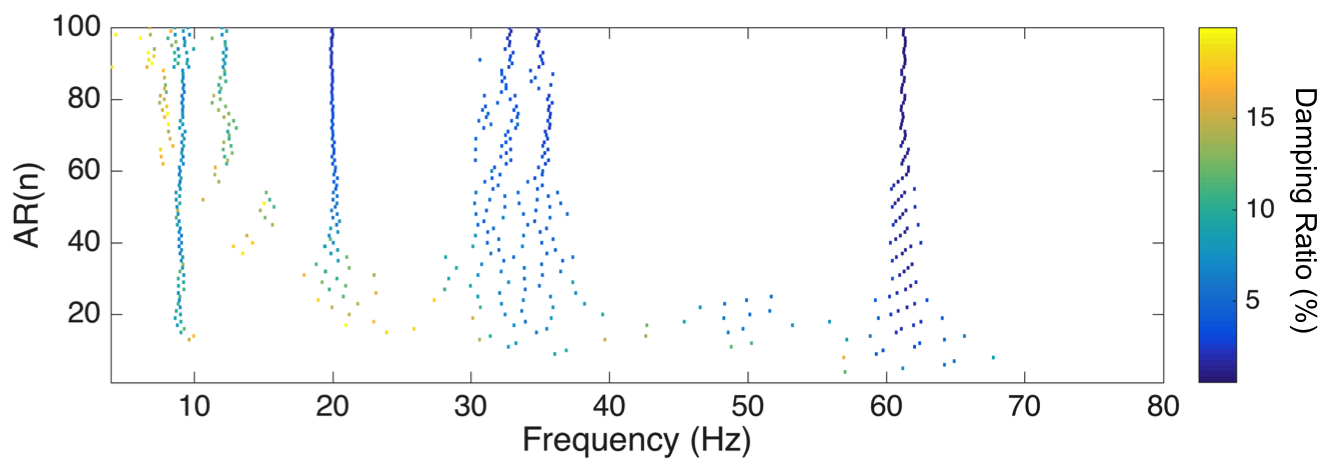}    
    \caption{\label{fig:stabplot} \textbf{AR stabilization diagram.} Stabilization diagram for AR$(n)$ type parametric models.}
\end{figure}

The modeling strategy consists of the successive fitting of AR$(n)$ models (with $n$ designating the AR order) until a suitable model is selected. Model parameter estimation is achieved by minimizing a quadratic prediction error (PE) criterion leading to a least squares (LS) estimator \cite[p. 206]{Ljung99}. Model order selection, which is crucial for successful identification, may be based on a combination of tools, including the Bayesian information criterion (BIC)  (Figure \ref{fig:BIC-RSS}a), which is a statistical criterion that penalizes model complexity (order) as a counteraction to a decreasing quality criterion \cite[pp. 505--507]{Ljung99}, monitoring of the RSS/SSS (residual sum of squares/ signal sum of squares) criterion (Figure \ref{fig:BIC-RSS}b), monitoring of the residual autocorrelation function (MATLAB function autocorr.m) \cite[p. 512]{Ljung99}, and use of ``stabilization diagrams'' (Figure \ref{fig:stabplot}) which depict the estimated modal parameters (usually frequencies) as a function of increasing model order \cite{Ljung99, Soderstrom-Stoica89}.


\begin{table}[b!]
\caption{Piezoelectric signal pre-processing for the parametric identification.} \label{tab:param-signals} 
\centering
\begin{tabular}{ll}
\hline 
Sampling frequency: & $f_s = 200$ Hz (after filtering and sub-sampling)  \\ 
Final bandwidth: & $[0.1 - 100]$ Hz  \\
Digital filtering: & Low-pass Chebyshev Type II  \\
Signal length: & $N = 4,000$ samples ($20$ s)  \\
\hline
\end{tabular}
\end{table}

An approximate plateau in the BIC and RSS/SSS sequences is achieved for model order $n > 50$  (Figure \ref{fig:BIC-RSS}), while the BIC value is minimized for order $n=72$. The AR$(72)$ model exhibits a very low RSS/SSS value of $0.7 \ \%$ demonstrating the accurate identification and successful dynamics representation by the specific model. Furthermore, as indicated by the frequency stabilization diagram of Figure \ref{fig:stabplot}, model orders of $n > 60$ are adequate for most natural frequencies to stabilize. Notice the vertical color bar in Figure \ref{fig:stabplot}, which presents the damping ratios for each estimated frequency for increasing model order. It may be observed that for the specific data set used in the baseline modeling process higher damping ratios are found within the $0.5 - 15$ Hz range.

The above identification procedure leads to an AR$(72)$ model. This model is used as reference and for providing approximate orders for the identification of the global VFP-AR model of the next section.

\subsection{Global identification under multiple flight states} \label{sec:VFP-results}

The parametric VFP-based identification of the wing dynamics is based on signals collected from the piezoelectric sensors under a series of wind tunnel experiments shown in Table \ref{tab:experiments}. The global modeling of the composite wing is based on signals obtained from a total of $M_1 \times M_2 = 144$ experiments. Airspeeds up to 17 m/s and AoA up to 15 degrees were considered for the VFP-based modeling procedure. The airspeed and AoA increments are $\delta k^1 = 1$ m/s and $\delta k^2 = 1$ degree, respectively, covering the corresponding intervals of $[9, 17]$ m/s and $[0, 15]$ degrees.
 
\begin{figure}[t!]
    \centering
    \includegraphics[scale=0.145]{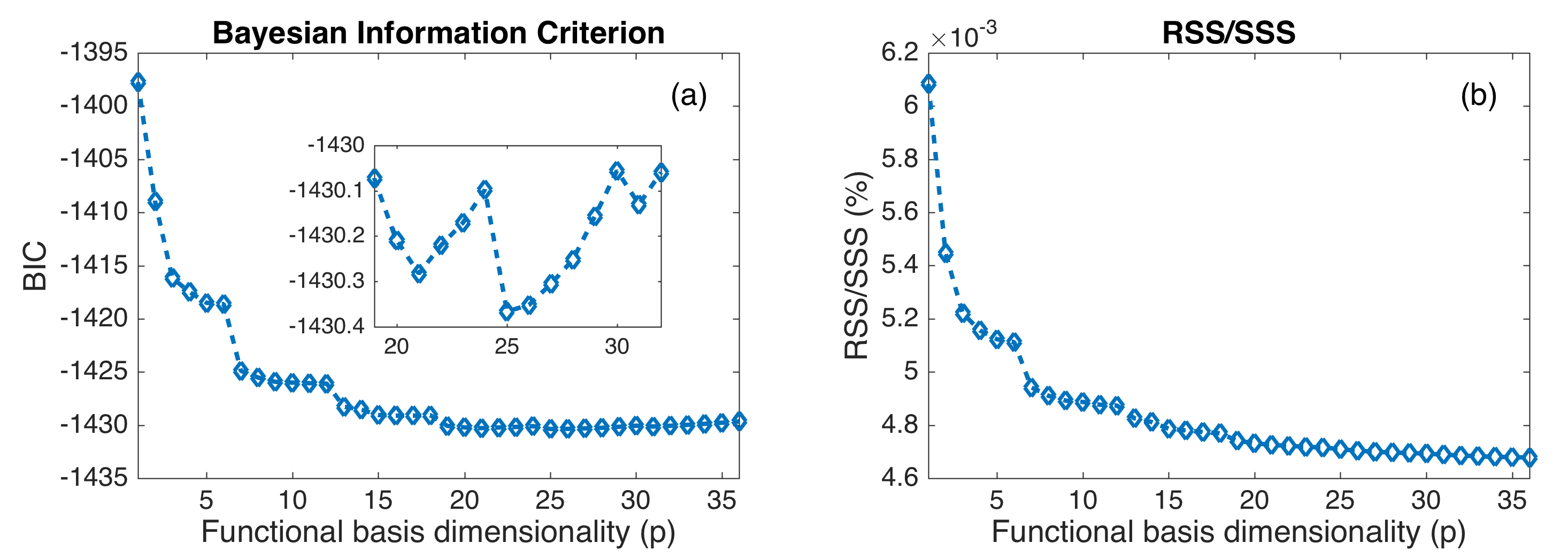}    
    \caption{\label{fig:VFP-criteria} \textbf{VFP model structure selection criteria.} Functional basis dimensionality selection for VFP-AR$(72)$ model and increasing number of basis functions $p$ (Chebyshev Type II bivariate polynomials): (a) BIC and (b) RSS/SSS.}
\end{figure}

Model order selection starts with the same order selected for the conventional AR models representing the wing dynamics for a constant indicative flight state. The final model order $n=72$ presently selected is based on the process outlined in the previous subsection and appropriate model validation techniques, such as checking the whiteness (uncorrelatedness) and the normality of the model residuals (MATLAB functions \textit{acf.m} and \textit{normplot.m}, respectively) \cite{Ljung99}. The functional subspace is selected via a similar approach based on the BIC and RSS/SSS criteria for increasing functional basis dimensionality (Figure \ref{fig:VFP-criteria}). An extended functional subspace consisting of $36$ Chebyshev Type II bivariate polynomial basis functions\cite{Kopsaftopoulos12, Dunkl-Xu01} is initially considered with the optimal functional basis subset selected based on minimization of the BIC criterion \cite{Kopsaftopoulos12}. From Figure \ref{fig:VFP-criteria}a it is evident that the minimum value of the BIC corresponds to a basis dimensionality of $p=25$ functions. Figure \ref{fig:VFP-criteria}b indicates that the RSS/SSS value for the selected functional subspace is in the order of $4.7 \cdot 10^{-3} \ \%$ indicating the extremely accurate representation of the wing dynamics by the VFP model. 

\begin{figure}[t!]
    \centering
    \includegraphics[scale=0.5]{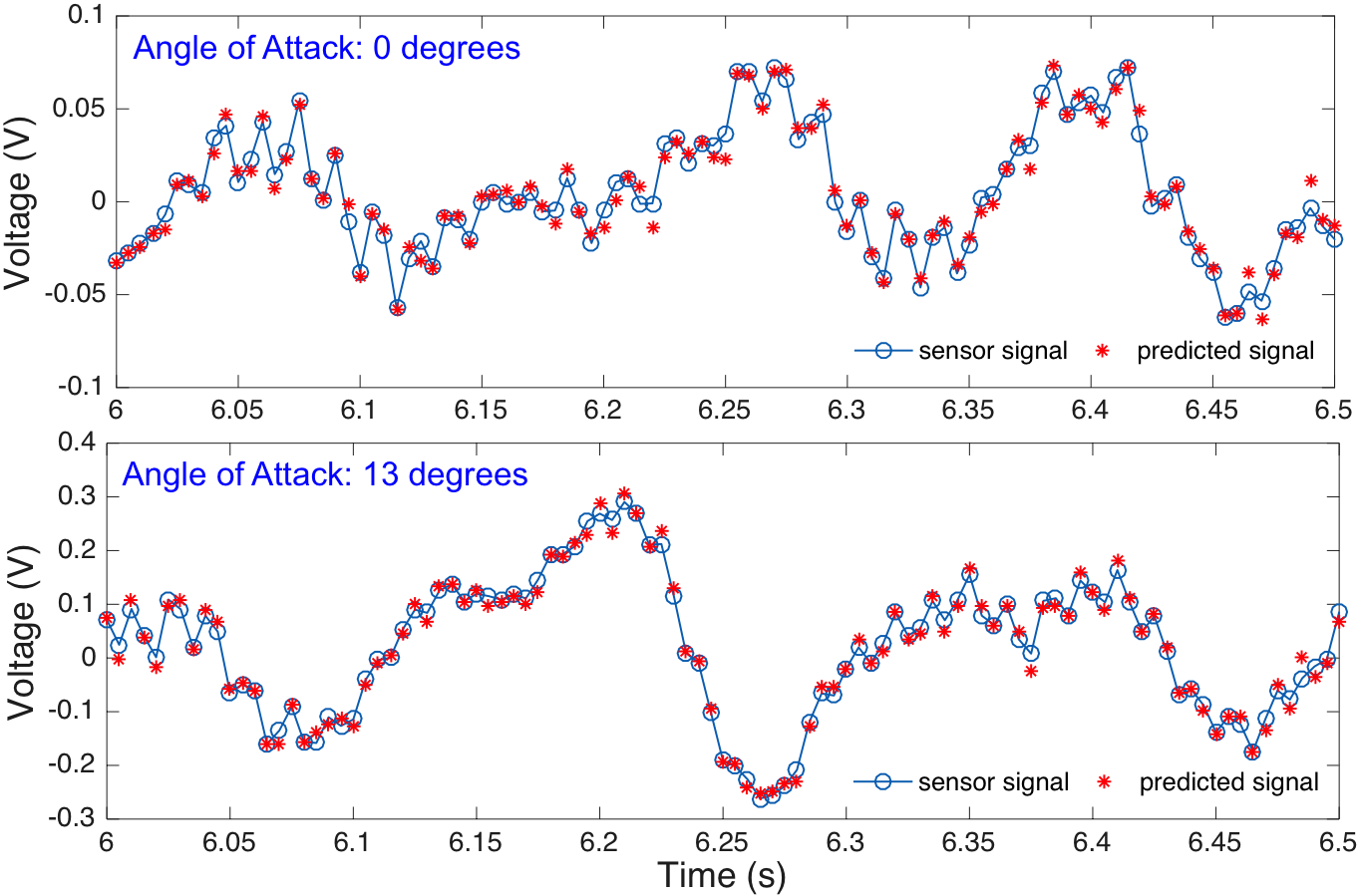}    
    \caption{\label{fig:VFP-predict} \textbf{VFP model predictions.} Functional basis dimensionality selection for VFP-AR$(72)$ model and increasing number of basis functions $p$ (Chebyshev Type II bivariate polynomials): (a) BIC and (b) RSS/SSS.}
\end{figure}

It is worth mentioning that the VFP-based RSS/SSS estimation takes into account all the residual series from all the data sets corresponding to the 266 different flight states that are used in the identification process. It may be observed that the RSS/SSS value of the global VFP model is significantly lower when compared to the corresponding RSS/SSS value of the baseline AR modeling stage ($4.7 \cdot 10^{-3} \ \%$ for the VFP versus $0.7 \ \%$ for the conventional AR model). This is to be expected, as one of the major advantages of the VFP model structure when compared to the LPV and multi-model approaches (see Section \ref{sec:intro}) is that it takes into account the data cross-correlation between all the sets that are used in the model identification process. This additional information, that is neglected in the LPV and multi-model modeling approaches, results in significantly improved parameter estimation accuracy and is reflected in the lower estimated variance of the residuals sequences. Hence, the VFP model identification stage results in a VFP-AR$(72)_{25}$ model. 

The predictive capability of the selected VFP-AR$(72)_{25}$ model is presented via indicative one-step-ahead prediction results for set airspeed of 15 m/s and AoA 0 (top plot) and 13 (bottom plot) degrees in Figure \ref{fig:VFP-predict}. The recorded signal points are shown as red circles ({\Red $o$}), while the VFP-mode-based predictions are depicted as blue asterisks ({\Blue$*$}). In both cases, the VFP model shows remarkable predictive capabilities, a fact that is also demonstrated by the very low RSS/SSS value (see Figure \ref{fig:VFP-criteria}).

\begin{figure}[t!]
    \centering
    \includegraphics[scale=1.65]{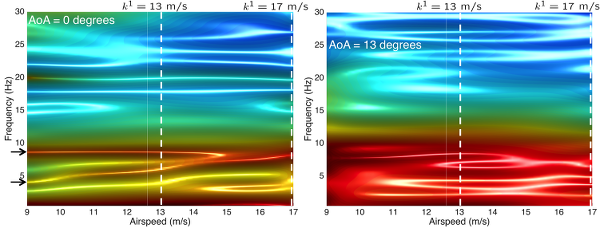}  
    \caption{\label{fig:vfp-frf-as} \textbf{Parametric VFP-based FRF magnitude versus airspeed.} Indicative parametric FRF magnitude results based on the VFP-AR$(72)_{25}$ global model for set AoA $k^2 = 0$ (left) and $k^2 = 13$ (right) degrees). The FRF magnitude is depicted as a function of frequency and airspeed. The two dashed vertical lines indicate the airspeeds $k^1 = 13$ m/s and $k^1 = 17$ m/s that correspond to the plots of Figure \ref{fig:vfp-frf-aoa}.}
\end{figure}

Indicative VFP-model-based frequency response function (FRF) magnitude results obtained from the VFP-AR$(72)_{25}$ global model are depicted as functions of frequency and airspeed for set AoA $k^2 = 0$ and $k^2 = 13$ degrees in Figure \ref{fig:vfp-frf-as}. The frequency resolution is $0.01$ Hz, while the airspeed resolution is $0.1$ m/s. The desired resolution can be completely defined based on the identified analytical functional dependence of the flight state vector with the model parameters and the corresponding functional subspaces (see Equation \ref{eq:vfp-ar-coef}). In the case of AoA $k^2 = 0$ degrees (left plot) observe how the wing mode at $4.5$ Hz for airspeed 9 m/s gradually increases with the increasing airspeed until completely coupled with the mode at $8.5$ Hz at approximately 16 m/s (the two modes are indicated with horizontal arrows). This behavior of  aeroelastic modes of the wing, as identified by the VFP-AR model, corresponds to the generation of the dynamic flutter phenomenon. It may be readily observed that the results of the left plot in Figure \ref{fig:vfp-frf-as} (AoA $k^2 = 0$ degrees) are, as expected, extremely accurate when compared to the corresponding non-parametric Welch-based analysis of Figure \ref{fig:PSDs-AS}. It is also worth mentioning that the non-parametric results of Figure \ref{fig:PSDs-AS} have been obtained using a significantly longer signal of 90 seconds, whereas the VFP-based parametric results are based on 20-seconds-long signals. 

The right plot of Figure \ref{fig:vfp-frf-as} presents the VFP-model-based FRF magnitude results for AoA $k^2 = 13$ degrees, which lies within the critical AoA stall range of the wing (see also Figure \ref{fig:xfoil}). It may be readily observed that the FRF magnitude (red indicates higher FRF magnitude, whereas blue indicates lower magnitude) is much higher than that of $k^2 = 0$ degrees (left plot) in the frequency range $[0.1 - 13]$ Hz due to the high-amplitude wing vibrations that are generated during the stall phenomenon. Furthermore, it may be also observed that the aeroelastic dynamic behavior of the wing in this frequency range it is more complicated: for airspeeds lower than 12 m/s there are two dominant aeroelastic modes; between 12 m/s and 15 m/s an additional mode shows up that is coupled with the mode at 8 Hz for airspeeds higher than 15 m/s. This complex dynamic behavior for AoA 13 degrees is caused by the simultaneous occurrence of different aeroelastic phenomena, such as stall and flutter and potential corresponding non-linearities (LCOs) \cite{Henshaw-etal07}. Via the use of such a global VFP model structure it is possible to enable aircraft control schemes in order to suppress, minimize, and even predict dynamic flutter via appropriate real-time monitoring techniques.

\begin{figure}[t!]
    \centering 
    \includegraphics[scale=1.65]{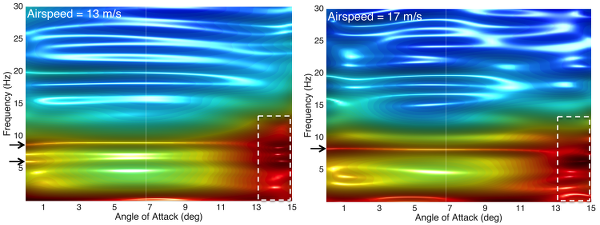}    
    \caption{\label{fig:vfp-frf-aoa} \textbf{Parametric VFP-based FRF magnitude versus AoA.} Indicative parametric FRF magnitude results based on the VFP-AR$(72)_{25}$ global model for set airspeeds $k^1 = 13$ m/s (left) and $k^1 = 17$ m/s (right). The FRF magnitude is depicted as a function of frequency and AoA. The dashed rectangular areas indicate the critical AoA stall range. }
\end{figure}

The vertical dashed lines in Figure \ref{fig:vfp-frf-as} indicate airspeeds $k^1 = 13$ m/s and $k^1 = 17$ m/s for which the VFP-model-based FRF magnitude curves obtained via the VFP-AR$(72)_{25}$ global model are depicted as functions of frequency and AoA in Figure \ref{fig:vfp-frf-aoa}. The frequency resolution is $0.01$ Hz, while the AoA resolution is $0.1$ degrees. The airspeed of 13 m/s (left plot in Figure \ref{fig:vfp-frf-aoa}) is before the occurrence of flutter and the wing exhibits two distinct aeroelastic modes, indicated with horizontal arrows, within the $[5 -10]$ Hz range (compare with the corresponding cross-section defined by the vertical dashed lines in Figure \ref{fig:vfp-frf-as}). On the other hand, for the airspeed of 17 m/s (right plot in Figure \ref{fig:vfp-frf-aoa}) the aforementioned modes have been coupled due to the existence of flutter (compare with Figure \ref{fig:vfp-frf-as}) and a single aeroelastic mode at $8.5$ Hz is dominant.

In addition, by observing the frequency evolution versus the AoA it may be assessed that the amplitude of the VFP-based FRF magnitude increases for lower frequencies ($< 15$ HZ) with increasing AoA as the wing approaches stall. More specifically, the FRF magnitude exhibits a sharp increase for AoA higher than 13 degrees in which stall occurs (compare with Figure \ref{fig:xfoil}). This is evident in both plots of Figure \ref{fig:vfp-frf-aoa}, $k^1 = 13$ m/s (left) and $k^1 = 17$ m/s (right), and the critical AoA stall range is indicated within the dashed rectangular areas. Also, in the case of 17 m/s (right plot in  Figure \ref{fig:vfp-frf-aoa}) it may be also observed the occurrence of complex dynamics for AoA higher than 13 degrees, in which both stall, flutter and corresponding non-linearities take place.

By comparing the VFP-based parametric FRF magnitudes with the corresponding non-parametric Welch-based spectral estimates of Figure \ref{fig:PSDs-AoA} it may be concluded that high accuracy is achieved by the global modeling approach which also employs a significantly shorter signal length (see Table \ref{tab:signal-details}).

\begin{figure}[t!]
    \centering
    \includegraphics[scale=0.45]{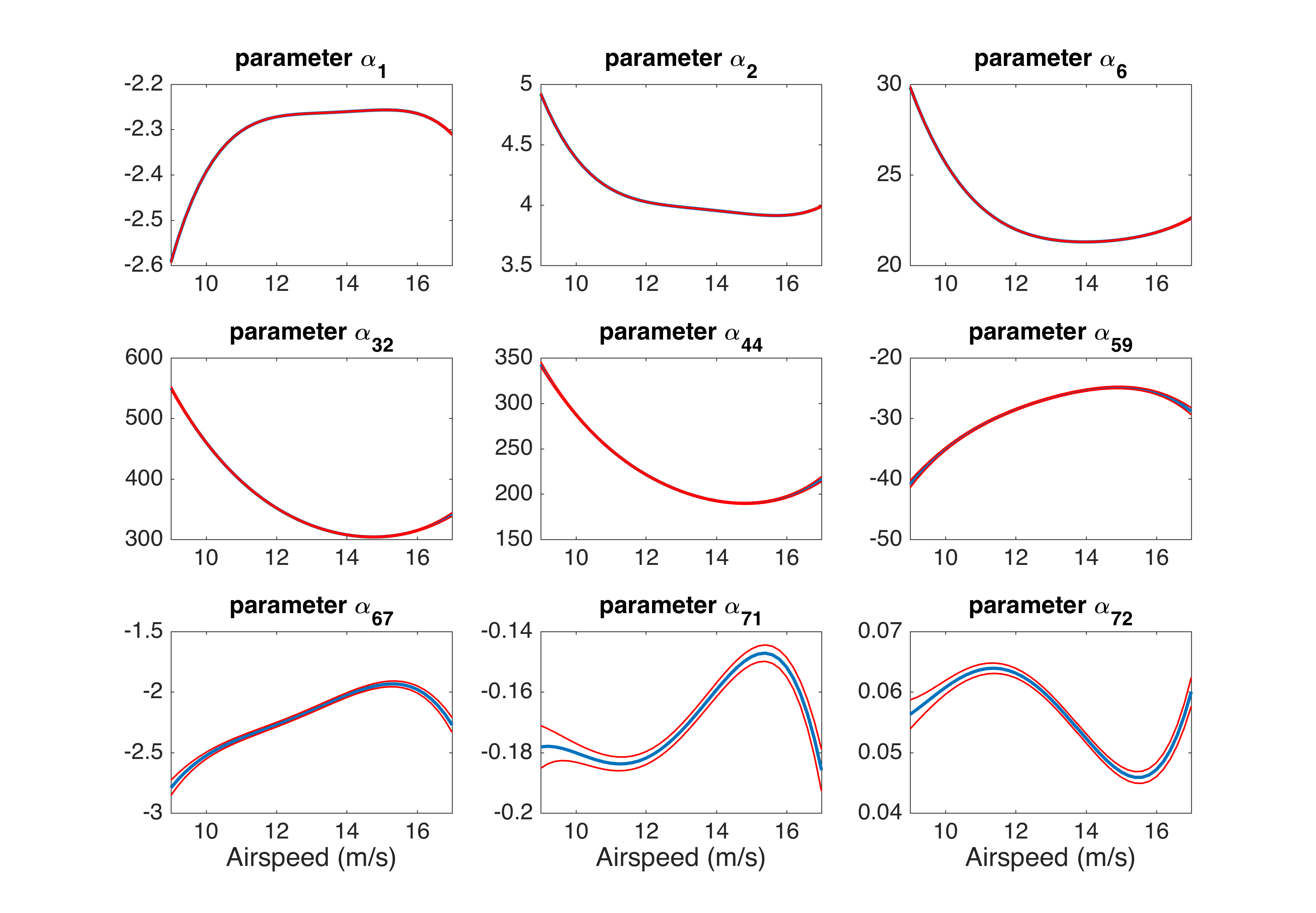}
    \caption{\label{fig:2D-model-param-AS} \textbf{2D VFP model parameters versus airspeed.} Indicative VFP-AR$(72)_{25}$ model parameters along with their $99\%$ confidence intervals versus airspeed for set AoA of $k^2 = 6$ degrees. }
\end{figure}
\begin{figure}[t!]
    \centering
    \includegraphics[scale=0.45]{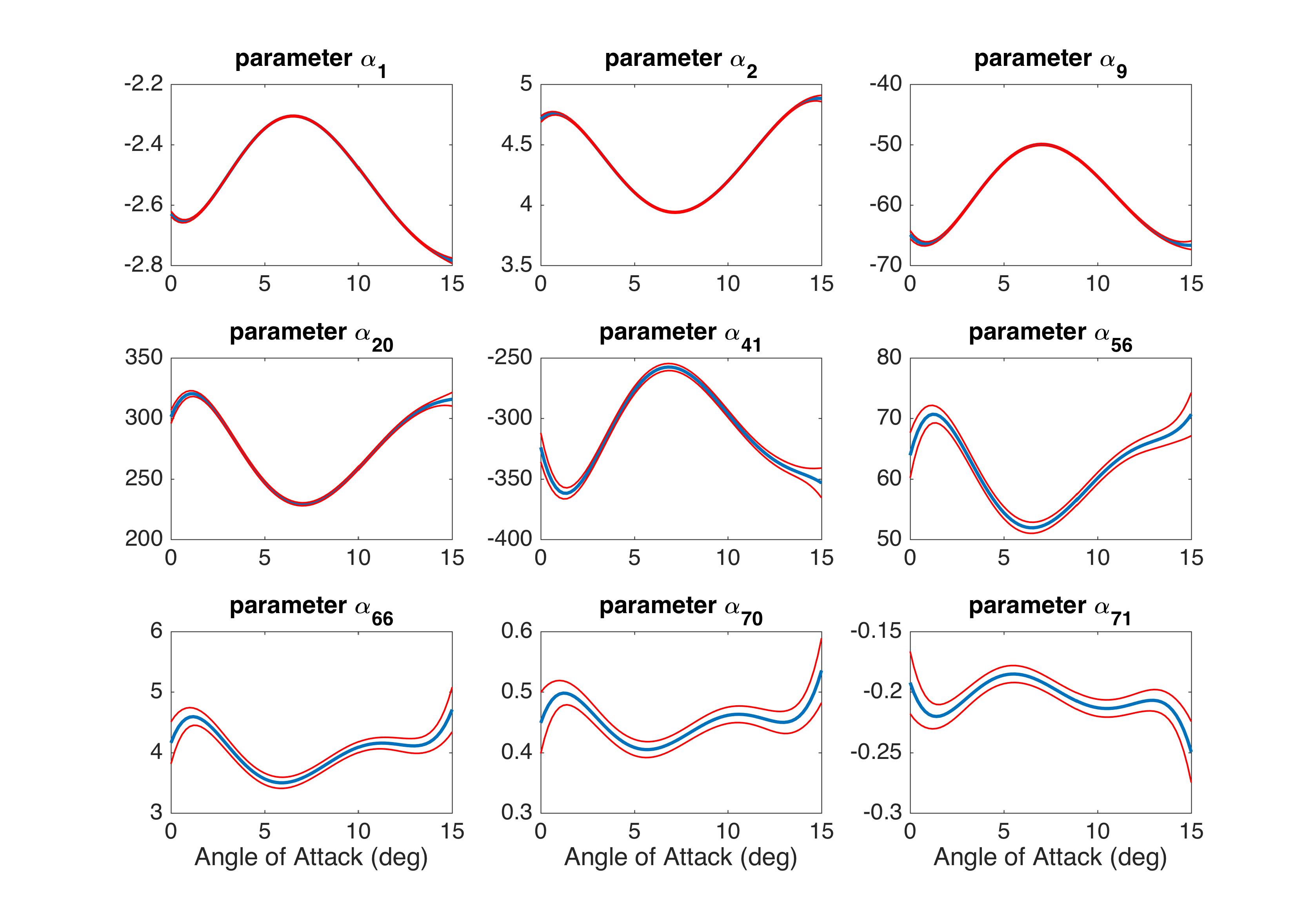}
    \caption{\label{fig:2D-model-param-AoA} \textbf{2D VFP model parameters versus AoA.} Indicative VFP-AR$(72)_{25}$ model parameters along with their $99\%$ confidence intervals versus AoA for set airspeed $k^1 = 17$ m/s.}
\end{figure}

Indicative 2-dimensional (2D) AR model parameters of the VFP-AR$(72)_{25}$ model as functions of the airspeed and AoA are depicted in Figures \ref{fig:2D-model-param-AS} and \ref{fig:2D-model-param-AoA}, respectively. The corresponding $99\%$ confidence intervals are also shown in red lines. In most of the cases. as shown in the various subplots, it may be readily observed that the confidence intervals are extremely narrow, which demonstrates the accuracy of the parameter estimation approach. In cases of increased uncertainty due to as reflected in the recorded signals, the stochastic identification approach will compensate by increasing the parameter estimation uncertainty and hence, leading to increased parameter confidence intervals.

\begin{figure}[t!]
    \centering
    \includegraphics[scale=0.5]{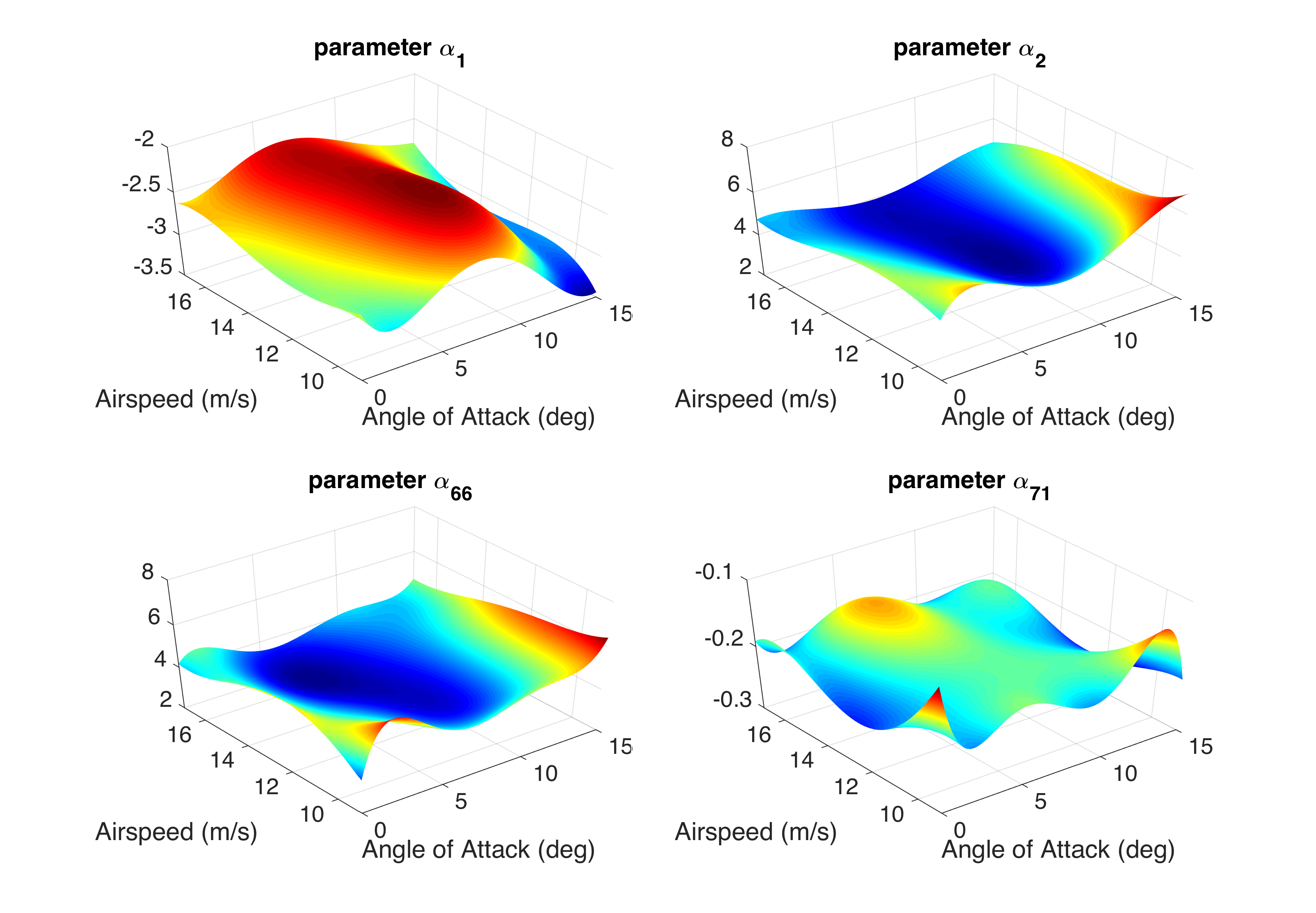}
    \caption{\label{fig:3D-model-param} \textbf{3D VFP model parameters versus airspeed and AoA.} Indicative VFP-AR$(72)_{25}$ model parameters as functions of both airspeed and AoA.}
\end{figure}

The VFP model parameters (Equation \ref{eq:vfp-ar-coef}) are projected into functional subspaces spanned by the selected basis functions consisting of bivariate polynomials parametrized in terms of airspeed and AoA. Therefore, the VFP model parameters constitute explicit functions of both the airspeed and AoA. Figure \ref{fig:3D-model-param} presents indicative 3D VFP-AR$(72)_{25}$ model parameters as functions of both the airspeed and AoA. This is an alternative representation of Figures \ref{fig:2D-model-param-AS} and \ref{fig:2D-model-param-AoA} showing the variation of the model parameters with respect to the varying flight states of the wing characterized by multiple airspeeds and AoA. 

\section{Concluding Remarks} \label{sec:conclusions}

The objective of this work was to introduce a novel data-based stochastic ``global'' identification framework for  flight and aeroelastic state awareness of aerospace vehicles. The proposed framework is based on the novel class of stochastic functionally pooled models for representing the system dynamics under varying flight states and uncertainty. In the context of aeroelastic state awareness, the authors introduced for the first time the use of Vector-dependent Functionally Pooled (VFP) models characterized by explicit functional dependencies between the flight states, model parameters, and the model residual sequences. The class of VFP models resembles the form of LPV models, with some critical differences: (i) the signals are treated as a single entity and potential cross-correlations are accounted for, (ii) the number of estimated parameters is minimal, (iii) and the estimation is accomplished in a single step (instead of two subsequent steps) for achieving optimal accuracy. 

For the experimental assessment and evaluation of the proposed stochastic framework, a prototype intelligent composite UAV wing was designed and fabricated at Stanford University. The composite wing was outfitted with bio-inspired distributed networks consisting of 148 micro-sensors embedded inside the composite layup. A series of wind tunnel experiments was conducted under various airspeeds and AoA for collecting data under multiple flight states (multiple equilibria points). A total of 266 wind tunnel experiments covering the complete range of the considered conditions was conducted. The postulated data-based stochastic identification approach that is based on the novel VFP time-series model structure achieved the accurate representation of the wing dynamics and aeroelastic behavior for all the admissible flight states and enabled the monitoring and detection of the dynamic stall and flutter phenomena. The obtained results demonstrated the effectiveness and accuracy of the stochastic ``global'' identification framework as a first step towards the next generation of ``fly-by-feel'' aerospace vehicles with state-sensing and awareness capabilities.

Current and future work addresses:
\begin{itemize}
\item Real-time extension and implementation of the proposed identification framework for on-board state awareness.
\item Extension of the global VFP models to account for fast evolving non-stationary dynamic behavior that is critical for a number of aerospace structural systems.
\item Integration with high-fidelity structural and aeroelastic computational models for increased physical insight, additional data generation under varying flight states and structural conditions for training purposes, and complete structural awareness from the material to the vehicle-wide level.
\item Extension of the developed framework to the case of multivariate global VFP models to simultaneously account for large numbers of sensors and both spatial and time data cross-correlation.
\item Postulation of appropriate control schemes for flutter suppression and mitigation, early stall detection and avoidance, gust alleviation, and optimized maneuvering and aeroelastic performance based on global models.
\end{itemize}

\section*{Acknowledgment}

This research was supported by the U.S. Air Force Office of Scientific Research (AFOSR) Multidisciplinary University Research Initiative (MURI) program under grant FA9550-09-1-0677 with Program Manager Byung-Lip (Les) Lee. The authors would like to thank Mr. Pengchuan Wang, Mr. Ravi Gondaliya, Dr. Jun Wu and Dr. Shaobo Liu for their help during the wind tunnel experiments. Finally, the authors would like to acknowledge the support of Dr. Lester Su and Prof. John Eaton in the wind tunnel facility at Stanford University.


\bibliographystyle{ieeetr}


\bibliography{wing_references} 

\end{document}